\documentclass[sn-mathphys,Numbered,dvipsnames]{sn-jnl}

\usepackage{graphicx}%
\usepackage{multirow}%
\usepackage{amsmath,amssymb,amsfonts}%
\usepackage{amsthm}%
\usepackage{mathrsfs}%
\usepackage[title]{appendix}%
\usepackage[table]{xcolor}%
\usepackage{textcomp}%
\usepackage{manyfoot}%
\usepackage{booktabs}%
\usepackage{algorithmicx}%
\usepackage{algpseudocode}%
\usepackage{listings}%

\usepackage[T1]{fontenc}
\usepackage[utf8]{inputenc}
\usepackage{amsmath}
\usepackage{amssymb}
\usepackage{colonequals}
    \usepackage{listings}
\usepackage[export]{adjustbox}
\usepackage{epstopdf}
\usepackage{subcaption}
\usepackage[]{algorithm,algpseudocode}

\usepackage{acro}
\usepackage{xspace}
\usepackage{needspace}
\usepackage[inline]{enumitem}
\usepackage{amsthm}

\usepackage{tikz}
\usetikzlibrary{positioning,arrows,backgrounds,automata,shapes,decorations.shapes,decorations.pathmorphing,decorations.markings,decorations.text,fit,patterns}
\tikzset{
  initial text=$ $, 
  >=stealth',
  every picture/.style=thick
}

\usepackage{pgfplots}
\pgfplotsset{compat=1.18}
\usepackage{multicol}

\newtheorem{definition}{Definition}
\newtheorem{example}{Example}
\newtheorem{theorem}{Theorem}
\newtheorem{lemma}{Lemma}
\newtheorem{remark}{Remark}

\let\IDeclareAcronym\DeclareAcronym
\renewcommand{\DeclareAcronym}[2]{%
 \IDeclareAcronym{#1}{%
  #2,foreign-plural={} 
  }
}
\DeclareAcronym{bdd}{
  short = BDD,
  long = Binary Decision Diagram
}

\DeclareAcronym{dbm}{
  short = DBM,
  long = Difference Bound Matrix,
  long-plural-form = Difference Bound Matrices
}

\DeclareAcronym{fpga}{
  short = FPGA,
  long = Field Programmable Gate Array,
  short-indefinite = an
}

\DeclareAcronym{ltl}{
  short = LTL,
  long  = Linear Temporal Logic,
  short-indefinite = an
}
\DeclareAcronym{ltl3}{
  short = LTL$_3$,
  long  = Three-value \ac{ltl},
  short-indefinite = an
}

\DeclareAcronym{mitl}{
  short = MITL,
  long = Metric Interval Temporal Logic,
  short-indefinite = an
}
\DeclareAcronym{msl}{
  short = MSL,
  long = Mars Science Laboratory,
  short-indefinite = an
}
\DeclareAcronym{mtl}{
  short = MTL,
  long = Metric Temporal Logic,
  short-indefinite = an
}
\DeclareAcronym{mfotl}{
  short = MFOTL,
  long = Metric First-Order Temporal Logic,
  short-indefinite = an
}

\DeclareAcronym{rv}{
  short = RV,
  long = Runtime Verification,
  short-indefinite = an
}
\DeclareAcronym{smc}{
  short = SMC,
  long = Statistical Model Checking,
  short-indefinite = an
}
\DeclareAcronym{stl}{
  short = STL,
  long = Signal Temporal Logic,
  short-indefinite = an
}
\DeclareAcronym{ta}{
  short = TA,
  long  = Timed Automaton,
  long-plural-form = Timed Automata
}
\DeclareAcronym{tba}{
  short = TBA,
  long  = Timed B\"uchi Automaton,
  long-plural-form = Timed B\"uchi Automata
}
\DeclareAcronym{tltl}{
  short = TLTL,
  long  = Timed \ac{ltl}
}
\DeclareAcronym{tre}{
  short = TRE,
  long = Timed Regular Expression
}

\DeclareAcronym{xml}{
  short = XML,
  long  = Extensible Markup Language,
  short-indefinite = an
}

\newcommand{\myquot}[1]{``#1''}

\newcommand{\set}[1]{\{#1\}}

\newcommand{\true}{\mathit{true}\xspace}
\newcommand{\false}{\mathit{false}\xspace}

\newcommand{\unknown}{\textbf{\textit{?}}}

\newcommand{\etal}{et al.\xspace}

\renewcommand{\mid}{:}

\newcommand{\realty}{\mathbb{R}}
\newcommand{\natty}{\mathbb{N}}
\newcommand{\ratty}{\mathbb{Q}}
\newcommand{\nnratty}{\mathbb{Q}_{\geq 0}}

\newcommand{\boolthreety}{\mathbb{B}_{3}}
\newcommand{\truthvaluegeq}{\succeq}

\newcommand{\nnreals}{\clockty}

\newcommand{\nnrats}{\ratty_{\geq 0}}

\newcommand{\powerset}[1]{2^{#1}}
\newcommand*{\Cdot}{\raisebox{-0.25ex}{\scalebox{1.6}{.}}}
\newcommand{\where}[1][1]{\Cdot\ifnum #1=0{}\else{\foreach \n in {1,...,#1}{\ }}\fi} 

\newcommand{\concat}{\cdot}

\newcommand{\clockty}{\realty_{\geq 0}}

\makeatletter
\newcommand{\pto}{}
\newcommand{\pgets}{}

\DeclareRobustCommand{\pto}{\mathrel{\mathpalette\p@to@gets\to}}
\DeclareRobustCommand{\pgets}{\mathrel{\mathpalette\p@to@gets\gets}}

\newcommand{\p@to@gets}[2]{%
  \ooalign{\hidewidth$\m@th#1\mapstochar\mkern5mu$\hidewidth\cr$\m@th#1\to$\cr}%
}
\makeatother

\renewcommand{\bar}[1]{\overline{#1}}

\newcommand{\infstates}[1]{\textit{inf}\,(#1)}
\newcommand{\empset}{\varnothing}
\newcommand{\twty}{T\Sigma}
\newcommand{\twdivty}{T\!\!\vcenter{\hbox{{\tiny{D}}}}\Sigma}
\newcommand{\twsymty}{T\!\vcenter{\hbox{{\tiny{S}}}}\Sigma}
\newcommand{\lang}{\phi}
\newcommand{\tbaty}{\mathbb{A}}
\newcommand{\automaton}{\mathcal{A}}
\newcommand{\compautomaton}{\overline{\mathcal{A}}}
\newcommand{\transition}[1]{\xrightarrow{(\sigma_{#1},t_{#1})}}
\newcommand{\transitiontau}[1]{\xrightarrow{(\sigma_{#1},\tau_{#1})}}
\newcommand{\transitiont}[1]{\xrightarrow{(\sigma_{#1},t_{#1}+t)}}
\newcommand{\transitiontm}[1]{\xrightarrow{(\sigma_{#1},t_{#1}-t)}}
\newcommand{\timelogic}{\ac{mitl}\xspace}
\newcommand{\itimelogic}{\iac{mitl}\xspace}
\newcommand{\languageof}[1]{\mathcal{L}(#1)}

\newcommand{\evalfunc}{\mathcal{V}\xspace}
\newcommand{\evaluate}[2]{\evalfunc(#1)(#2)}
\newcommand{\evalfuncdiv}{\mathcal{V}_D\xspace}
\newcommand{\evaluatediv}[2]{\evalfuncdiv(#1)(#2)}
\newcommand{\monitorfunc}{\mathcal{M}\xspace}
\newcommand{\monitor}[2]{\monitorfunc(#1)(#2)}

\newcommand{\tbadiv}{\automaton_{\textit{D}}}
\newcommand{\nonempty}[1]{S_{#1}^{\textit{ne}}}
\newcommand{\terminal}[2]{\mathcal{T}_{#1}(#2)}

\newcommand{\tbacap}{\otimes}
\newcommand{\clocks}{\mathcal{X}}

\makeatletter
\newcommand{\shortsim}{%
  \settowidth{\@tempdima}{n}
  \resizebox{\@tempdima}{\height}{$\sim$}%
}
\makeatother

\newcommand{\monitaal}{\textsc{MoniTAal}\xspace}
\newcommand{\mitppl}{\textsc{MightyPPL}\xspace}
\newcommand{\monpoly}{\textsc{MonPoly}\xspace}
\newcommand{\reelay}{Reelay\xspace}
\newcommand{\timescales}{{Timescales}\xspace}

\makeatletter
\newcommand{\shorteq}{%
  \settowidth{\@tempdima}{n}
  \resizebox{\@tempdima}{\height}{=}%
}
\makeatother

\usepackage[color=Purple!40]{todonotes}

\begin{document}

\title[Efficient Monitoring of Timed Properties]{Efficient Monitoring of Timed Properties}


\author*[1]{\fnm{Thomas Møller} \sur{Grosen}}\email{tmgr@cs.aau.dk}

\author[2]{\fnm{Sean} \sur{Kauffman}}\email{sean.k@queensu.ca}

\author[1]{\fnm{Kim Guldstrand} \sur{Larsen}}\email{kgl@cs.aau.dk}

\author[1]{\fnm{Martin} \sur{Zimmermann}}\email{mzi@cs.aau.dk}


\affil[1]{\orgname{Aalborg University}, \orgaddress{\city{Aalborg}, \country{Denmark}}}

\affil[2]{\orgname{Queens University}, \orgaddress{\city{Kingston}, \state{Ontario}, \country{Canada}}}

\abstract{In this paper we study monitoring of real-time systems with respect to properties given by a pair of Timed Büchi Automata, one for the property and one for its complement. This includes properties expressible in temporal logics that are closed under complementation and can be translated into Timed Büchi Automata, e.g., Metric Interval Temporal Logic.

We introduce efficient symbolic online monitoring algorithms in a number of settings, using difference bound matrices representing zones. Our contributions include a principled treatment of time divergence and monitoring under timing uncertainty. Our online monitoring procedure is implemented in the tool \monitaal, and shown to effectively monitor properties over long traces.}




\maketitle

\section{Introduction}
\label{sec:introduction}

Runtime monitoring has gained acceptance as a method for formally verifying the correctness of executing systems.
Monitoring means to test a sequence of observations of a system against a specification, often written in a formal logic.
Monitoring contrasts with static verification methods, like model checking, in that it is computationally easier due to only testing a single system execution.
Runtime monitoring may also be applied to black-box systems where details about the environment and design of the monitored system are not required to be known in advance.

Many systems have so-called ``extra-functional'' requirements that must be expressed with respect to time.
These systems, generally called real-time systems, are pervasive in modern life as the controllers of cyber-physical systems.
To express requirements with time components, logics such as \ac{mtl} and derivatives like \ac{mitl} have been developed that extend the more classical \ac{ltl} with timing constraints~\cite{koymans1990mtl,alur1996mitl}.
Finite Automata have also been extended with time to form Timed Automata~\cite{alur1994tba}.
These formalisms allow the expression of notions such as that ``a response shall occur within 20 milliseconds of a request.''

Monitoring timed properties is possible and several solutions have been proposed, each with their own advantages and drawbacks.
In this work, we introduce an efficient solution to the online monitoring problem for timed properties under time divergence.
Given a property and a finite timed sequence, our method determines if the property is guaranteed to be satisfied or violated by any continuation of that finite sequence.
Additionally, the eventual satisfaction or violation of a property considers that any future timing constraints will eventually be settled.
Note that online monitoring here contrasts with offline monitoring, where the system is assumed to have terminated and timed properties are interpreted with finite semantics.
In online monitoring, liveness properties (e.g., ``eventually, a request shall be received'') cannot be violated since there will always be more symbols, while they can be violated in offline monitoring, since the entire sequence is known.

In this paper we study the monitoring of real-time properties expressed by a pair of Timed Büchi Automata (TBA), one for the property and one for its complement.
Note that this setting includes properties given in Metric Interval Temporal Logic (\timelogic), as formulas of \timelogic can be translated into equivalent TBA~\cite{brihaye2017mightyl} under a point-wise semantics. 
We consider properties over infinite timed words where time is given with infinite precision and its passage is represented by real numbers.
Our symbolic approach exploits so-called zones\footnote{Also known as DBMs: Difference Bound Matrices.}, which are used for efficient model checking of Timed Automata \cite{DBLP:conf/ac/BengtssonY03}. 

In the first part, we offer a new, simple way of dealing with time divergence. Time divergence means that, during the infinite behavior of a real-time system, time progresses beyond any finite bound. Time divergence is stated as an assumption by most works in the area because it reflects how time behaves in reality. However, the algorithmic support for time divergence in earlier work on monitoring seems somewhat underdeveloped.

To understand how time divergence impacts monitor verdicts, consider the property ``nothing should be observed after an hour.''
It should be clear that time divergence guarantees that no infinite sequence will satisfy this property.
Because we are interested in online monitoring of properties over infinite timed sequences, the language of the property is empty, and its monitor should evaluate any finite prefix to be in violation of it.
Conversely, a monitor that does not compensate for time divergence will register an \emph{unknown} verdict for finite prefixes that do not include observations past the hour mark, as time can converge before an hour has passed.

Then, we deal with timing uncertainty, i.e., a setting where the real-valued time-points of events can only be observed up to a given precision. 
We show that our symbolic monitoring algorithm can easily be adapted to handle timing uncertainty.
Finally, we report on a prototype implementation of the presented algorithm and present experiments based on an industrial case study. We also integrate the formula translation tool \mitppl~\cite{hkmmp2025b} to run benchmarks from \timescales~\cite{DBLP:conf/rv/Ulus19} in order to compare with the state of the art monitoring tools \monpoly~\cite{basin2011monpoly} and \reelay~\cite{ulus2026OnlineMonitoringMetric}.

This work is an extended version of an invited contribution presented at FORMATS~2022~\cite{GKLZ24}. Here, we additionally present more detailed explanations, full proofs, and a prototype implementation.

\section{Related Work}
\label{sec:related}

Many techniques to monitor timed properties have focused on monitoring logics with finite-word semantics.
The first work to introduce timed property monitoring is by Roşu~\etal focusing on discrete-time finite-word \ac{mtl}~\cite{rosu2005monitoring}.
Basin~\etal proposed algorithms for monitoring real-time finite-word properties in~\cite{basin2012algorithms} and compared the differences between different time models.
Ulus~\etal described monitoring \acp{tre} for finite words using unions of two-dimensional zones~\cite{ulus2014offline,ulus2016online}.

The most closely related work to ours is that by Bauer~\etal in which the authors introduced the classical \ac{ltl3} monitor construction and then showed how a similar construction can be used for \ac{tltl}~\cite{bauer2006monitoring}.
Their method transforms \iac{tltl} formula to event-clock automata which are strictly less expressive than \acp{tba}, which we support.
Their algorithm also differs from ours in being based on the so-called region automata.  
The region construction was introduced as a method to partition the infinite state space of \iac{tba} using an equivalence relation between clock assignments~\cite{alur1994tba}.
In the region construction, two clock valuations are equivalent if they agree on the integer parts of the clock values and the order of the fractional parts.
This forms a finite representation of the infinite state space, but the number of regions is exponential in the parameters of the automaton~\cite{alur1994tba}.
Despite this complexity, more compact representations are possible in many cases and can often be achieved using so-called \emph{zones}~\cite{yannakakis1997EfficientAlgorithmMinimizing}.
In theory, regions can be easier than zones to work with because a region graph is canonical (whereas zones must be normalized) and all valuations within a region are bisimilar.
However, in many cases, zone-based algorithms provide an order of magnitude performance improvement over region-based algorithms (see, e.g.,~\cite{DBLP:conf/fct/LarsenPY95}).


Two more recent works have proposed solutions to the problem of monitoring timed languages specified in \ac{mtl}.
Baldor~\etal showed how to construct a monitor for a dense-time \ac{mitl} formula by constructing a tree of timed transducers~\cite{baldor2013monitoring}.
They showed how subsets of \ac{mitl} could be used to limit the complexity of their technique which requires linear space in the size of the input for the full fragment.
Ho~\etal split unbounded and bounded parts of a dense-time \ac{mitl} formula for monitoring, using traditional \ac{ltl} monitoring for the unbounded parts and permitting a simpler construction for the (finite-word) bounded parts~\cite{ho2014online}.
Unlike the work by Baldor~\etal, their method is size independent of the input.
However, it does require non-elementary blowup of the formula to ensure no unbounded operators appear in the context of a bounded operator.
They also monitor bounded parts using a dynamic programming formulation that relies on a maximum bound for the number of events in a time span.
Neither the solution by Baldor~\etal or Ho~\etal address time-divergence.

Although the previously mentioned works do not implement their solutions, some tools have been released for monitoring \ac{mtl}-based logics.
Basin~\etal implemented \monpoly, which can monitor an expressive finite-word (safety) fragment of \ac{mfotl} using discrete-time semantics~\cite{basin2008runtime}.
Bulychev~\etal implemented a rewrite-based monitoring algorithm similar to the one proposed by Roşu~\etal~\cite{rosu2005monitoring} for Weighted \ac{mtl} in the \textsc{Uppaal} \ac{smc} tool~\cite{bulychev2013rewrite}.
R2U2 is a tool for generating monitors for \acp{fpga} developed by Moosbrugger~\etal that supports finite-word \ac{mitl} properties~\cite{moosbrugger2017r2u2}.
More recently, Chattopadhyay and Mamouras presented a verified monitor for discrete, past-time (finite word) \ac{mitl} with quantitative semantics~\cite{chattopadhyay2020verified}.
Ulus introduced a method to monitor past-time (finite word) \ac{mtl} safety properties over a rational time domain using sequential networks~\cite{ulus2026OnlineMonitoringMetric} and implemented it in the tool \reelay, which also supports quantified properties.

There is also work on monitoring for other, more expressive timed logics.
Monitoring tools for \acp{tre} and similar logics are limited to discrete-time such as the AERIAL tool by Basin~\etal to monitor metric regular languages using rewriting~\cite{basinAERIALAlmostEventRate} and Montre by Ulus that monitors \acp{tre} using zones~\cite{ulus2017MontreToolMonitoring}.
\Ac{stl} adds the complexity of signals to \ac{mitl}.
Nikovic and Maler introduced AMT to monitor \ac{stl} for past and bounded dense-time properties under a bounded variability assumption~\cite{nickovic2007AMTPropertyBasedMonitoring}.
Nikovic and Yamaguchi later extended this work in their RTAMT tool to support \ac{stl} robustness semantics and add new interfaces~\cite{nickovicRTAMTOnlineRobustness2020}.
The monitoring algorithms of those tools exploit their bounded-time and bounded-variability assumptions to support conversion to deterministic \ac{ta}~\cite{henzinger1992WhatGoodAre}.
Havelund and Peled added support to a timed version of their DejaVu tool to monitor a past-time \ac{tltl} with quantification over data.
They decompose properties into BDDs for each subformula~\cite{havelund2020FirstOrderTimedRuntime}.
Henry~\etal recently introduced an algorithm to monitor dense-time properties over approximate timed words in a distributed setting, which bears some resemblance to our work on timing uncertainty~\cite{henry2025DistributedMonitoringTimed}. However, they restrict the properties they monitor to those representable by deterministic \ac{ta}, which substantially simplifies the problem.
Some tools also exist to convert timed logics to automata which we will cover in the next section.

The topic of monitorability has recently been studied in the timed setting~\cite{moni, amara:hal-05043055}. Monitorability of a property refers to whether or not the property can effectively be monitored i.e., the possibility of deciding satisfaction or violation from a finite prefix.

After the publication of the conference version~\cite{GKLZ24} of this work, the monitoring approach presented here has been extended to a setting with delayed observations~\cite{FGLZ24} and a setting with assumptions and incomplete observations~\cite{CGLTZ24}.

\section{Preliminaries}
\label{sec:prelim}

We first define some notation used throughout the paper.
The set of natural numbers (excluding zero) is $\natty$, we define~$\natty_0 = \natty \cup \{0\}$, the set of rational numbers is $\ratty$, the set of non-negative rational numbers  is $\nnratty$, 
the set of real numbers is $\realty$, and the set of non-negative real numbers  is $\clockty$. 
The three-valued set of monitor verdicts is $\boolthreety = \{\top, \unknown, \bot\}$. 
Given a set $S$ the set of all its subsets is denoted $\powerset{S}$.
The cross product of two sets $S$ and $T$ is $S\times T$.
Given a sequence $\sigma$, $\sigma_i$ denotes the element at the $i$th position of $\sigma$ (where one is the first position) and $\sigma^i$ denotes the suffix of $\sigma$ starting at index $i$.
Given two sequences $s$ and $t$, we write $s \concat t$ to denote their concatenation.

A timed word over a finite alphabet $\Sigma$ is a pair $\rho = (\sigma, \tau)$ where $\sigma$ is an untimed word over $\Sigma$ and and $\tau$ is a sequence of non-decreasing non-negative real numbers of the same length as $\sigma$.
Timed words may be finite or infinite where the set of finite timed words is $\twty^*$ and the set of infinite timed words is $\twty^\omega$.
We also represent a timed word as a sequence~$(\sigma_1,\tau_1), (\sigma_2,\tau_2), \ldots$ of pairs of letters and non-negative reals.  If $\rho=(\sigma_1,\tau_1), (\sigma_2,\tau_2), \ldots (\sigma_n,\tau_n)$ is a finite timed word, we denote by $\tau(\rho)$ the total time duration of $\rho$, i.e., $\tau_n$ (with the convention~$\tau(\varepsilon) = 0$).

If $\rho_1=(\sigma^1_1,\tau^1_1),(\sigma^1_2,\tau^1_2),\ldots,(\sigma^1_n,\tau^1_n)$ is a finite timed word and $\rho_2=(\sigma^2_1,\tau^2_1),(\sigma^2_2,\tau^2_2),\ldots$ a finite or infinite timed word and $t \in \nnreals$ then the timed word concatenation $\rho_1 \cdot_t \rho_2$ is defined iff $t \ge \tau(\rho_1)$. Then, we define~$\rho_1 \cdot_t \rho_2 = (\sigma_1,\tau_1),(\sigma_2,\tau_2),\ldots$ such that 
\[\sigma_i = \begin{cases}
    \sigma^1_i & \text{if } i\le n,\\
    \sigma^2_{i-n} & \text{else,}
\end{cases}\quad\text{and}\quad
\tau_i = \begin{cases}
    \tau^1_i & \text{if } i\le n,\\
    \tau^2_{i-n}+t & \text{else}.
\end{cases}\]

\subsection{Metric Interval Temporal Logic}
We use the Metric Interval Temporal Logic, \timelogic, in this work to formalize examples because we can translate it into the \acp{tba} that we use in our monitoring algorithm.
Brihaye~\etal developed the tool MightyL to translate \ac{mitl} formulas to \acp{tba} in a compositional manner~\cite{brihaye2017mightyl}.
Some earlier work implemented algorithms to translate subsets of \ac{mitl} to \acp{tba} as well.
Li~\etal proposed and implemented Casaal, a tool to construct deterministic approximations of \acp{tba} from \ac{mtl}$_{0,\infty}$ formulas~\cite{li2017practical,bulychev2012monitor}.
Geilen and Dams implemented an algorithm to produce a deterministic \ac{ta} for dense-time \ac{mitl}$_{\leq}$ (a subset of \ac{mtl}$_{0,\infty}$) using an on-the-fly tableau construction that discretizes the time domain and only supports an upper bound~\cite{geilen2000tableau}.
Note that some other works exist that provide algorithms for the translation of \ac{mitl} or related logics to \acp{tba}, but without providing implementations~\cite{finkbeiner2009circuits,nickovic2010dta}.

Let $\Sigma$ be a finite alphabet. The syntax of \timelogic formulas over $\Sigma$ is given by the grammar
\[
\varphi \coloncolonequals p\ |\ \neg \varphi\ |\ \varphi \vee \varphi\ |\ X_I \varphi\ |\ \varphi\ U_{I} \varphi\,
\]
where $p \in \Sigma$ and $I$ is a non-singular interval over $\clockty$ with endpoints in ${\natty_0 \cup \{+\infty\}}$.
We often write $\shortsim\,n$ for $I=\{d\in\realty\mid d \shortsim n\}$ where $\shortsim \in \{<,\leq,\geq,>\}$, and $n \in \natty_0$.
Note that we consider a pointwise semantics, which is the reason we add the next modality to the syntax. This  differs from the original definition of \timelogic, which did not have a next modality, as it used a continuous semantics (cf.~\cite[Section~29.6]{handbook} for a detailed discussion about the differences between the semantics).

The semantics of \timelogic is defined over infinite timed words.
Given such a timed word $\rho = (\sigma_1,\tau_1),(\sigma_2,\tau_2),\ldots \in \twty^\omega$, a position $i \geq 1$, and \itimelogic formula $\varphi$, we inductively define the satisfaction relation $\rho, i \models \varphi$ as  follows:

\begin{center}
\begin{tabular}{p{.72cm}p{0.9cm}ll}
  ${\rho,i \models}$ & ${ p }$ & \textit{if} & ${p = \sigma_i}$ \\
  ${\rho,i \models}$ & ${ \neg \varphi}$ & \textit{if} & ${\rho,i \not\models \varphi}$ \\
  ${\rho,i \models}$ & ${ \varphi \vee \psi}$ & \textit{if} & ${\rho,i \models \varphi \textit{ or } \rho,i \models \psi}$ \\
  ${\rho,i \models}$ & ${ X_I \varphi}$ & \textit{if} & ${\rho,{(i+1)} \models \varphi} \textit{ and } \tau_{i+1} - \tau_i \in I$\\
  ${\rho,i \models}$ & ${ \varphi\ U_{I} \psi}$ & \textit{if} & ${\exists k \geq i \where \rho,k \models \psi, \tau_k - \tau_{i} \in I \textit{ and } \forall j \where i \leq j < k\where \rho,j \models \varphi}$\\
\end{tabular}
\end{center}
We write $\rho \models \varphi$ whenever $\rho, 1 \models \varphi$, and say that $\rho$ satisfies $\varphi$. 
Given \itimelogic formula~$\varphi$, its language $\languageof{\varphi}$ is the set of all infinite timed words that satisfy $\varphi$.

We also define the standard syntactic sugar
$\true = p \vee \neg p$,
$\false = \neg \true$,
$\varphi \wedge \psi = \neg (\neg \varphi \vee \neg \psi)$,
$\varphi \rightarrow \psi = \neg \varphi \vee \psi$,
$F_{I} \varphi = \true\ U_{I} \varphi$, 
and $G_{I} \varphi = \neg F_{I} \neg \varphi$.

\subsection{Timed Automata}

\Iac{tba} $\automaton$ is a six-tuple $(Q, Q_0, \Sigma, \clocks, \Delta, \mathcal{F})$, where $\Sigma$ is a finite alphabet, $Q$ is a finite set of locations, $Q_0 \subseteq Q$ is a set of initial locations, $\clocks$ is a finite set of clocks, $\Delta \subseteq Q \times Q \times \Sigma \times \powerset{\clocks} \times G(\clocks)$ is a finite set of transitions with $G(\clocks)$ being the sets of constraints over $\clocks$, and $\mathcal{F} \subseteq Q$ is a set of accepting locations.
A transition~$(q,q',\alpha,\lambda, g)$ is an edge from $q$ to $q'$ on input symbol $\alpha$ where $\lambda$ is the set of clocks to reset and $g$ is a clock constraint over $\clocks$.
A clock constraint is a conjunction of atomic constraints of the form $x \sim n$, where $x$ is a clock, $n \in \natty_0$, and $\sim\ \in \{<, \leq, =, \geq, >\}$.
A state of \iac{tba} is a pair $(q,v)$ where $q$ is a location in $Q$ and $v : \clocks \rightarrow \clockty$ is a valuation mapping clocks to their values.
We say that for any $d \in \clockty$, $v+d$ is the valuation where $d$ is added to all clock values in $v$.

A run of $\automaton$ from a state $(q_0, v_0)$ is a sequence of steps over an infinite timed word~$(\sigma, \tau) \in \twty^\omega$ of the form
\[(q_0, v_0) \transitiontau{1} (q_1,v_1) \transitiontau{2} (q_2,v_2) \transitiontau{3} \cdots \]
where for all $i \geq 1$ there is a transition $(q_{i-1},q_{i},\sigma_{i},\lambda_i,g_i)$ such that $v_{i}(x) = 0$ for all $x$ in $\lambda_i$ and $v_{i-1}(x) + (\tau_i - \tau_{i-1})$ otherwise, and $g$ is satisfied by the valuation $v_{i-1}+(\tau_{i} - \tau_{i-1})$, where we use $\tau_0 = 0$. 
Given a run $r$, we denote the set of locations visited infinitely many times by $r$ as $\infstates{r}$.
A run $r$ of $\automaton$ is accepting if $\infstates{r} \cap \mathcal{F} \neq \empset{}$.
The language of $\automaton$ from a starting state $(q,v)$, denoted $\languageof{\automaton,(q,v)}$, is the set of all timed words with an accepting run in $\mathcal{A}$ starting from $(q,v)$.
We define the language of $\automaton$, written $\languageof{\automaton}$, to be $\bigcup_q \languageof{\automaton,(q,v_0)}$, where $q$ ranges over all locations in $Q_0$ and where $v_0(x) = 0$ for all $x \in \clocks$.

\begin{theorem}[\cite{alur1996mitl,brihaye2017mightyl}]
\label{thm:mtltba}
For each \ac{mitl} formula $\varphi$ there exists a \ac{tba} $\automaton$ with $\languageof{\varphi} = \languageof{\automaton}$.
\end{theorem}

Given two \acp{tba} $\automaton = (Q, Q_0, \Sigma, \clocks, \Delta, \mathcal{F})$ and $\automaton' = ( Q', Q'_0, \Sigma, \clocks', \Delta', \mathcal{F}')$, their intersection is denoted $\automaton \tbacap \automaton' = ( Q^{\tbacap}, Q_0^{\tbacap}, \Sigma, \clocks^{\tbacap}, \Delta^{\tbacap}, \mathcal{F}^{\tbacap})$, where 
\begin{itemize}[topsep=0pt]
    \item $Q_{\tbacap} = Q \times Q' \times \{1,2\}$,
    \item $Q_0^{\tbacap} = Q_0 \times Q'_0 \times \{1\}$,
    \item $\clocks^{\tbacap} = \clocks \cup \clocks'$ (we assume they are disjoint),
    \item $\Delta^{\tbacap} = \Delta_1^{\tbacap} \cup \Delta_2^{\tbacap}$ with
    \begin{itemize}
        \item[] $\Delta_1^{\tbacap} = \{((q_1,q'_1,1),(q_2,q'_2,i),\alpha,\lambda \cup \lambda',g \wedge g') \mid (q_1,q_2,\alpha,\lambda,g) \in \Delta$ and $(q'_1,q'_2,\alpha,\lambda',g') \in \Delta'$ and $i = 2$ if $q_1 \in \mathcal{F}$ else $i = 1\}$ and
        \item[] $\Delta_2^{\tbacap} = \{((q_1,q'_1,2),(q_2,q'_2,i),\alpha,\lambda \cup \lambda',g \wedge g') \mid (q_1,q_2,\alpha,\lambda,g) \in \Delta$ and $(q'_1,q'_2,\alpha,\lambda',g') \in \Delta'$ and $i = 1$ if $q'_1 \in \mathcal{F}'$ else $i = 2\}$,
    \end{itemize}
    \item and $\mathcal{F}^{\tbacap} = Q \times \mathcal{F}' \times \{2\}$.
\end{itemize}

\begin{lemma}[\cite{alur1994tba}]\label{lemma:intersection}
$\languageof{\automaton \tbacap \automaton'}=\languageof{\automaton}\cap\languageof{\automaton'}$.
\end{lemma}

\section{Monitoring in a Timed Setting}
\label{sec:monitoring}

In this section, we describe monitoring and show how it applies in the timed setting.
To this end, we first, and briefly, introduce monitoring in the untimed case and then extend it to the timed case, following the approach of Bauer~et al.~\cite{bauer2006monitoring}.

Traditionally in \ac{rv}, properties are specified using a temporal logic such as \ac{ltl} and a monitor is constructed from those properties.
A monitor is a kind of program that takes a finite word as an input and returns a \emph{verdict} depending on the relationship between the input and the property from which the monitor is constructed.
Verdicts are usually of the form \emph{accept} ($\top$), \emph{reject} ($\bot$), or \emph{unknown} ($\unknown$), although larger verdict domains exist to provide more information.

In online monitoring (our interest), the properties specify behaviors over infinite sequences of symbols, or words, while the monitor must interpret those specifications over an ever-growing finite prefix of such an infinite word.
The most prevalent solution to this problem is to use a monitor semantics where acceptance or rejection means that the finite word \emph{determines} the property and no future suffix can alter the verdict.
In the case where the finite prefix does not determine the property, the monitor outputs an \emph{unknown} verdict and continues.

Monitoring languages of timed infinite words works in much the same way as in the untimed setting. 
A finite prefix of an infinite timed word is checked to see if it determines the property.
If all possible infinite extensions of the prefix result in a word that is included in the monitored property, then the monitor returns the $\top$ verdict.
If no possible infinite extensions lead to a word that is included in the monitored property, then the monitor returns the $\bot$ verdict.
If extensions exist that could lead to either outcome, then the monitor returns $\unknown$ and continues monitoring. 
In the timed setting, we also need to know at what time point the prefix ends and when the extension starts. Intuitively, this concatenation time point represents knowledge about the time that has passed since the last observation of the prefix.

\begin{definition}[Monitor verdicts for timed languages]
\label{def:monitor}
Given a language of infinite timed words $\lang \subseteq \twty^\omega$, a finite timed word $\rho \in \twty^*$, and a time-point~$t \in \nnreals$ such that $t \ge \tau(\rho)$, the function $\evalfunc : \powerset{\twty^\omega} \rightarrow (\twty^* \times \nnreals) \rightarrow \boolthreety$ evaluates to a verdict with the following definition:
\[
\evaluate{\lang}{\rho,t} = \left.
  \begin{cases}
    \top & \text{if $\rho \concat_t \mu \in \lang\ $ for all $ \mu \in \twty^{\omega}$}, \\
    \bot & \text{if $\rho \concat_t \mu \notin \lang\ $ for all $ \mu \in \twty^{\omega}$}, \\
    \unknown & \text{otherwise}.
  \end{cases}
  \right.
\]
If $t < \tau(\rho)$, then $\evaluate{\lang}{\rho,t}$ is undefined.
\end{definition}

\begin{example}
\label{ex:response}
Consider the bounded response property ``whenever $a$ is observed, $b$ should be observed within 30 time units'' that is specified by the \timelogic formula $\varphi = G_{\ge 0}(a \rightarrow F_{\leq 30} b)$ where $\Sigma = \{a,b,c\}$.
This property corresponds to the \ac{tba} shown in Figure~\ref{fig:tba-response}.
This type of bounded response property is very common for real-time systems \cite{gear}.
It states that some trigger $a$ is followed by a reaction $b$ before the deadline elapses.
Note that we draw the ``sink'' state $q_3$ here for illustrative purposes as we will return to this example later in the paper, but it can be omitted without changing the language of the automaton.

\begin{figure}[ht]
  \centering
  \begin{tikzpicture}[shorten >=1pt,node distance=2cm,on grid,auto, thick] 
  \node[state,accepting,initial] (q1)  {$q_1$}; 
  \node[state] (q2) [right=3cm of q1] {$q_2$};
  \node[state] (q3) [right=3cm of q2] {$q_3$};
    \path[->] 
    (q1) edge [loop above] node {$b, c$} ()
         edge [bend left=20] node[above,align=center] {$a$\\$x \colonequals 0$} (q2)
    (q2) edge [loop above] node[align=center] {$a,c$\\ $x \le 30$} ()
         edge [bend left=20] node[below,align=center] {$b$\\$x \le 30$} (q1)
         edge node[above,align=center] {$a,b,c$\\$x > 30$} (q3)
    (q3) edge [loop above] node {$a,b,c$} ();
  \end{tikzpicture}
  \caption{\Ac{tba} corresponding to $G_{\ge 0}(a \rightarrow F_{\leq 30} b)$}
  \label{fig:tba-response}
\end{figure}

Now consider the finite timed prefix $\rho = (b,10), (a, 20)$. The verdict in this case is
\[\evaluate{\languageof{\varphi}}{\rho, t} = 
\begin{cases}
  \unknown & \text{if } t \le 50,\\
  \bot & \text{otherwise.} \end{cases}\]
For $t \le 50$ there are infinite extensions of the prefix that satisfy the property and those that violate it, e.g., we have 
\[\rho \cdot_{40} (b,0),(b,10),(b,20),\ldots = (b,10), (a, 20)(b,40),(b,50),(b,60),\ldots \in \languageof{\varphi}\] and 
\[\rho \cdot_{40} (b,20),(b,30),(b,40),\ldots = (b,10), (a, 20)(b,60),(b,70),(b,80),\ldots \notin \languageof{\varphi}.\]
When $t > 50$, then there are no accepting extensions as the $a$ at time~$20$ has not been followed by a $b$ within $30$ time units. 

This property demonstrates a way in which timed monitoring differs from the untimed setting. 
If we remove the time constraint and consider the (unbounded) response property $G(a \rightarrow F b)$, no finite prefix could ever determine the property and so the only possible verdict would be $\unknown$.
This is a classic example of what is called an \emph{unmonitorable} property~\cite{pnueli2006psl,bauer2011runtime,kauffman2021what}.
\end{example}

\section{Time Divergence}
Note that Definition~\ref{def:monitor} does not take time divergence into account, i.e., a verdict can be based on convergent extensions of a given prefix. As we will see in Example~\ref{ex:gt20}, this leads to verdicts that are only based on Zeno behaviors. 
In this section, we consider the monitoring problem in the presence of  time divergence and show how it affects the verdicts from monitors for timed languages.
Time divergence entails that time will always progress beyond any given time-bound.
Let us stress here that our algorithms work both in a Zeno and non-Zeno setting. 

We begin by defining the set of infinite time-divergent words.
These are the only timed words that will occur in practice, since time always diverges.
Mathematically, however, the set~$\twty^\omega$ includes words that are not time divergent, e.g., $(\alpha,\frac{1}{2}),(\alpha,\frac{3}{4}),(\alpha,\frac{7}{8}),\ldots$.
The definition states that time-divergent words are those where the time sequence is unbounded.
Note that we do not consider \emph{finite} timed words either divergent or convergent even though their time sequences technically converge.

\begin{definition}
The set of all time-divergent words $\twdivty^\omega \subseteq \twty^\omega$ is the set of all timed words $(\sigma_1,\tau_1), (\sigma_2,\tau_2) \ldots$ such that $\lim_{i\to\infty} \tau_i = +\infty$.
\end{definition}

We now use the set of time-divergent words to define a verdict function that accounts for time divergence.
Crucially, the properties that we monitor may include non-time-divergent words.
In that case, the verdict returned by the evaluation function under time divergence $\evalfuncdiv$ may differ from the verdict returned by $\evalfunc$.

\begin{definition}[Monitor verdicts under time divergence]
\label{def:monitorwithdivergence}
Given a language of infinite timed words $\lang \subseteq \twty^\omega$, a finite timed word $\rho \in \twty^*$, and a time-point~$t \in \nnreals$ such that $t \ge \tau(\rho)$, the function $\evalfuncdiv : \powerset{\twty^\omega} \rightarrow (\twty^* \times \nnreals) \rightarrow \boolthreety$ evaluates to a verdict with the following definition:
\[
\evaluatediv{\lang}{\rho,t} = \left.
  \begin{cases}
    \top & \text{if $\rho \concat_t \mu \in \lang\ $ for all $ \mu \in \twdivty^{\omega}$}, \\
    \bot & \text{if $\rho \concat_t \mu \notin \lang\ $ for all $ \mu \in \twdivty^{\omega}$}, \\
    \unknown & \text{otherwise}.
  \end{cases}
  \right.
\]
If $t < \tau(\rho)$, then $\evaluatediv{\lang}{\rho,t}$ is undefined.
\end{definition}

\begin{example}
\label{ex:gt20}
Consider the property ``the system will continue past 20 time units after the first observation'' represented by the \timelogic formula $\varphi = F_{\geq 20} a$, where $\Sigma = \{a\}$. 
This property is captured by the \ac{tba} shown in Figure~\ref{fig:tba-gt20}.

\begin{figure}[ht]
  \centering
  \begin{tikzpicture}[shorten >=1pt,node distance=2cm,on grid,auto,thick] 
  \node[state, initial] (q0) [left=3cm of q1] {$q_1$};
  \node[state,initial] (q1)  {$q_2$}; 
  \node[state,accepting] (q2) [right=3cm of q1] {$q_3$};
    \path[->] 
    (q0) edge node[above,align=center] {$a$\\$x := 0$} (q1)
    (q1) edge [loop above] node[above,align=center] {$a$\\$x < 20$} ()
         edge node[above,align=center] {$a$\\$x \ge 20$} (q2)
    (q2) edge [loop above] node {$a$} ();
  \end{tikzpicture}
  \caption{\Ac{tba} corresponding to $F_{\geq 20} a$}
  \label{fig:tba-gt20}
\end{figure}

If this property is monitored, the verdict may change depending on whether time divergence is accounted for.
Under time divergence, this property is clearly a tautology since all infinite time-divergent words must eventually reach location~$q_3$.
However, if time divergence is not assumed as in Definition~\ref{def:monitor}, then it is possible for an infinite timed word to never pass time to $x \ge 20$ after the first observation, and therefore stay in location $q_2$.

Suppose, for example, the finite timed prefix $\rho = (a,0), (a,10)$.
Since $\varphi$ is a tautology under time divergence, $\evaluatediv{\languageof{\varphi}}{\rho, t} = \top$ for all time points~$t\ge 10$. 
However, $\evaluate{\languageof{\varphi}}{\rho, t} = \unknown$ for all $t < 20$, since $\twty^\omega$ contains time-convergent words~$\mu$ such that $\rho\cdot_t\mu$ is not in the language of $\varphi$, e.g., for $t = 10$ and $\mu = (a,9),(a,9.9),(a,9.99),(a,9.999),\ldots$ we have $\rho\cdot_t\mu = (a,0), (a,10),(a,19),(a,19.9),(a,19.99),(a,19.999),\ldots \notin \languageof{\varphi}$. 
\end{example}

To ensure that verdicts are correct for all properties, we monitor an intersection of the given automata with a special \ac{tba} that only accepts time-divergent words.
This \ac{tba}, which we will call $\tbadiv$, is shown in Figure~\ref{fig:tba-divergence}.
The automaton must visit the left location infinitely often to accept and it can only visit this location once time has passed a threshold of one time unit.
Note that the exact threshold is arbitrary and could be any number; the purpose is to ensure that the language of the automaton is exactly the language of time-divergent words.

\begin{figure}[ht]
  \centering
  \begin{tikzpicture}[shorten >=1pt,node distance=2cm,on grid,auto,thick] 
   \node[state,initial,accepting] (q1)  {$A$};
   \node[state] (q2) [right=3cm of q1] {$B$};
    \path[->] 
    (q1) edge [bend left=20] node[above,align=center] {$\Sigma$\\$z\colonequals 0$} (q2)
    (q2) edge [loop right] node[align=center] {$\Sigma$\\ $z < 1$} ()
         edge [bend left=20] node[below,align=center] {$\Sigma$\\$z \ge 1$} (q1);
  \end{tikzpicture}
  \caption{\Ac{tba} $\tbadiv$ capturing time divergence}
  \label{fig:tba-divergence}
\end{figure}

\begin{remark}
The language of $\tbadiv$ is exactly the set of all time-divergent words.
\end{remark}

We will use the automaton~$\tbadiv$ to restrict the possible extensions of an observation to divergent words.

\begin{example}
\label{ex:gt20comptd}
We now consider the complement property to Example~\ref{ex:gt20} that is captured by the \timelogic formula $\bar{\varphi} = G_{\geq 20} \false$.
Note that, since we use symbols in our \timelogic formulas and not propositions, we cannot write $\neg a$ here but must use its complement $\Sigma \setminus \{a\} = \varnothing$ which is equivalent to $\false$ in our \timelogic syntax.
The \ac{tba} for $\bar{\varphi}$ is shown in Figure~\ref{fig:tba-gt20comp}.

\begin{figure}[ht]
  \centering
  \begin{tikzpicture}[shorten >=1pt,node distance=2cm,on grid,auto,thick] 
  \node[state,initial] (q0) [left =3cm of q1] {$\bar{q_1}$};
  \node[state,initial,accepting] (q1)  {$\bar{q_2}$}; 
     \path[->]
     (q0) edge node[align=center] {$a$\\ $x := 0$} (q1)
     (q1) edge [loop above] node[align=center] {$a$\\ $x < 20$} ();
  \end{tikzpicture}
  \caption{\Ac{tba} corresponding to $G_{\geq 20} \false$}
  \label{fig:tba-gt20comp}
\end{figure}

Now we want to intersect this \ac{tba} with $\tbadiv$ to restrict its language to only time-divergent words.
The result of this operation is shown in Figure~\ref{fig:tba-gt20comptd}. Note that the intersection automaton contains the unreachable locations $(\bar{q_1}, A, 2)$, $(\bar{q_1}, B, 1)$, $(\bar{q_1}, B, 2)$,  and $(\bar{q_2}, A, 1)$ that we omit for simplicity.

\begin{figure}[ht]
  \centering
  \begin{tikzpicture}[shorten >=1pt,node distance=2cm,on grid,auto,thick]
  \node[state,initial,rectangle, rounded corners=10] (q2) [] {$(\bar{q_1},A,1)$};
  \node[state,rectangle, rounded corners=10] (q1) [right=5cm of q2] {$(\bar{q_2},B,1)$};
  \node[state,accepting,rectangle, rounded corners=10] (q3) [below=2.5cm of q1] {$(\bar{q_2},A,2)$};
  \node[state,rectangle, rounded corners=10] (q4) [below=2.5cm of q2] {$(\bar{q_2},B,2)$};

    \path[->]
    (q1) edge[bend right = 15] node[near start,yshift=.55cm,above,left,align=center] {$a$\\ $x<20$\\ $z< 1$} (q4)
    (q1) edge[bend right] node[align=center,left] {$a$\\ $x < 20$\\ $z \ge 1$} (q3)
    (q2) edge node[left,align=center] {$a$\\ $x\colonequals 0$\\ $z\colonequals 0$} (q4)
    (q3) edge[bend right] node[right,align=center] {$a$\\ $x < 20$\\$z\colonequals 0$} (q1)
    (q4) edge [loop left] node[align=center] {$a$\\ $ x < 20$\\$z<1$} ()
         edge node[below,align=center] {$a$\\ $x < 20$\\ $z \ge 1$} (q3);
  \end{tikzpicture}
  \caption{\Ac{tba} for the intersection of the \ac{tba} from Figure~\ref{fig:tba-gt20comp} (corresponding to $G_{\geq 20} \false$) and $\tbadiv$ from Figure~\ref{fig:tba-divergence} (state names are as described in the construction above Lemma~\ref{lemma:intersection}). Unreachable locations have been omitted.}
  \label{fig:tba-gt20comptd}
\end{figure}

In this \ac{tba}, an accepting run must visit (the only accepting) location $(\bar{q_2}, A, 2)$ infinitely often.
The only transition leaving $(\bar{q_2}, A, 2)$ resets the clock $z$ and all transitions leading to $(\bar{q_2}, A, 2)$ have the guard~$z \ge 1$.
Hence, at least one unit of time has to pass between any two visits to $(\bar{q_2}, A, 2)$.
However, the clock~$x$ is never reset and each transition has the guard~$x < 20$. 
Hence, there cannot be an accepting run and the product automaton has an empty language.
\end{example}

\section{A Symbolic Method for Monitoring}
\label{sec:solution}

In this section, we describe an algorithm to monitor languages of infinite timed words that correctly accounts for time divergence.
Our algorithm is loosely based on the classical constructions for monitoring \ac{ltl3} and \ac{tltl} by Bauer~\etal~\cite{bauer2006monitoring}, but with alterations to address the many differences between the timed and untimed domains.

Our solution requires that properties are specified as two \acp{tba} - one for the property and one for its complement.
Although \acp{tba} are not closed under complementation, we consider this requirement to not be too much of a limitation.
This is because we expect a property to be expressed by a user in \timelogic, which, along with its negation, can be converted to \acp{tba} using one of the methods described in Section~\ref{sec:prelim}. 

Before we address the monitoring algorithm, we first introduce ``reach-sets'' and ``non-empty language states'' of \iac{tba}.
Given \iac{tba} $\automaton$, a state~$(q,v)$ has a non-empty language when it has an accepting run starting in $(q,v)$, i.e., if $\languageof{\automaton, (q,v)} \neq \varnothing $.

\begin{definition}\label{def:non-empty}
Given \iac{tba} $\automaton = (Q, Q_0, \Sigma, \clocks, \Delta, \mathcal{F})$, the set of states with non-empty language is $\nonempty{\automaton} = \{ (q,v) \mid q \in Q, v \in \clocks \rightarrow \clockty \where \languageof{\automaton, (q,v)} \neq \varnothing \}$.
\end{definition}

In the following definition, we write $(q_0, v_0) \xrightarrow{\rho}_\automaton (q_n, v_n)$ for a finite timed word~$\rho = (\sigma, \tau) \in \twty^*$ to denote the existence of a finite sequence 
\[(q_0, v_0) \transitiontau{1} (q_1,v_1) \transitiontau{2} \cdots \transitiontau{n} (q_n,v_n) \]
of states where for all $1 \leq i \leq n$ there is a transition $(q_{i-1},q_{i},\sigma_{i},\lambda_i,g_i)$ such that $v_{i}(x) = 0$ for all $x$ in $\lambda_i$ and $v_{i-1}(x) + (\tau_i - \tau_{i-1})$ otherwise, and $g$ is satisfied by the valuation $v_{i-1}+(\tau_{i} - \tau_{i-1})$, where we use $\tau_0 = 0$. 
Note that this definition is analogous to the definition of infinite runs in Section~\ref{sec:prelim}. 

\begin{definition}\label{def:state-estimate}
Given \iac{tba} $\automaton$, a finite timed word $\rho \in \twty^*$, and a time-point~$t \in \nnreals$ with $t \ge \tau(\rho)$, the reach-set of $\automaton$ over $\rho$ at time~$t$, the set of possible states a run over $\rho$ starting from initial states of $\automaton$ can end in after time~$t$ has passed, is given by
\begin{align*}
    \terminal{\automaton}{\rho,t} = \{(q, v + (t-\tau(\rho))) \mid (q_0, v_0) \xrightarrow{\rho}_\automaton (q, v)\\ \text{ where } q_0 \in Q_0 \text{ and }v_0(x) = 0 \text{ for all } x \in \clocks\}.
\end{align*}
\end{definition}

\begin{remark}
    The reach-set $\terminal{\automaton}{\rho,t}$ is always a finite set of states of $\automaton$.
\end{remark}

We can now define a function for a monitor verdict using Definitions~\ref{def:non-empty}~and~\ref{def:state-estimate}.
To determine the verdict for a language $\phi \subseteq \twty^\omega$, the function requires both \iac{tba} $\automaton$ such that $\languageof{\automaton} = \phi$ and its complement, $\compautomaton$.
Definition~\ref{def:monitorsymbolic} states that a finite timed word $\rho \in \twty^*$ positively (negatively) determines the property under time divergence if the states of $\compautomaton \tbacap \tbadiv$ ($\automaton\tbacap\tbadiv$) with non-empty languages are disjoint from the reach-set of $\compautomaton \tbacap \tbadiv$ ($\automaton\tbacap\tbadiv$) over $\rho$.

\begin{definition}[Monitoring \acp{tba} under time divergence]\label{def:monitorsymbolic}
Given \iac{tba}~$\automaton$, its complement $\compautomaton$ ($\languageof{\compautomaton} = \twty^\omega \setminus \languageof{\automaton}$), the automaton $\tbadiv$ such that $\languageof{\tbadiv} = \twdivty^\omega$, a finite timed word $\rho \in \twty^*$, and a time-point~$t \in \nnreals$ such that $t \ge \tau(\rho)$, the function~${\monitorfunc : \tbaty \times \tbaty \rightarrow (\twty^* \times \nnreals) \rightarrow \boolthreety}$ (where $\tbaty$ denotes the set of \acp{tba}) computes a verdict with the following definition:
\[
\monitor{\automaton, \compautomaton}{\rho, t} = \left.
  \begin{cases}
    \top & \text{if } \terminal{\compautomaton \tbacap \tbadiv}{\rho,t} \cap \nonempty{\compautomaton \tbacap \tbadiv} = \varnothing, \\
    \bot & \text{if } \terminal{\automaton \tbacap \tbadiv}{\rho,t} \cap \nonempty{\automaton \tbacap \tbadiv} = \varnothing, \\
    \unknown & \text{otherwise.}
  \end{cases}
  \right.
\]
If $t < \tau(\rho)$, then $\monitor{\automaton, \compautomaton}{\rho, t}$ is undefined.
\end{definition}

The following theorem shows that the symbolic definition of $\monitorfunc$ coincides with the definition of $\evalfuncdiv$ in Definition~\ref{def:monitorwithdivergence}.

\begin{theorem}
\label{thm:correctness}
$\monitor{\automaton, \compautomaton}{\rho,t} = \evaluatediv{\languageof{\automaton}}{\rho,t}$ for all $\rho \in \twty^*$ and all $t \in \nnreals$ such that $t \ge \tau(\rho)$.
\end{theorem}

\begin{proof}
Let $\automaton'$ be an arbitrary \ac{tba}. We will show that $\terminal{\automaton'}{\rho,t} \cap \nonempty{\automaton'} $ is nonempty iff there exists a $\mu \in \twty^\omega$ (not necessarily time divergent) with $\rho \cdot_{t}\mu \in \languageof{\automaton'}$.
Then, we obtain 
\begin{itemize}
    \item $\monitor{\automaton,\compautomaton}{\rho,t} = \top$ iff $\evaluatediv{\languageof{\automaton}}{\rho,t} = \top$ by instantiating the equivalence for $\automaton' = \compautomaton\otimes\tbadiv$, and 
    \item $\monitor{\automaton,\compautomaton}{\rho,t} = \bot$ iff $\evaluatediv{\languageof{\automaton}}{\rho,t} = \bot$ by instantiating the equivalence for $\automaton' = \automaton\otimes\tbadiv$.
\end{itemize}
This completes the proof, as both functions only have three elements in their codomain and we have shown that two of them have the same preimage  w.r.t.\ both functions.

So, let $\terminal{\automaton'}{\rho,t} \cap \nonempty{\automaton'} \neq\emptyset$. Then, by definition, there is a state~$(q,v')$ of $\automaton'$ such that
\begin{itemize}
    \item $(q_0, v_0) \xrightarrow{\rho}_{\automaton'} (q, v)$ for some $q_0 \in Q_0$ and for $v_0$ with $v_0(x) =0$ for all $x$ such that $v' = v + (t - \tau(\rho))$, and
    \item  there is an accepting infinite run of $\automaton'$ starting in $(q,v')$ that processes some $\mu \in \twty^\omega$.
\end{itemize}
Let $\rho = (\sigma_1, t_1) \cdots (\sigma_n,t_n)$, which satisfies $t_n = \tau(\rho) \le t$.
Then, due to $(q_0, v_0) \xrightarrow{\rho}_{\automaton'} (q, v)$, there is a run prefix
\[(q_0, v_0) \transition{1} (q_1,v_1) \transition{2} \cdots \transition{n} (q,v). \]
Also, let $\mu = (\sigma_{n+1}, t_{n+1}) (\sigma_{n+2},t_{n+2})\cdots$ and let 
\[
(q, v') \transition{n+1} (q_{n+1}, v_{n+1}) \transition{n+2} (q_{n+2}, v_{n+2}) \transition{n+3}\cdots
\]
be the accepting run $\automaton'$ on $\mu$. 

These two runs can be combined into the run
\begin{align*}
&(q_0, v_0) \transition{1} (q_1,v_1) \transition{2} \cdots \transition{n} (q,v) 
\transitiont{n+1} \\
&(q_{n+1}, v_{n+1}) \transitiont{n+2} (q_{n+2}, v_{n+2}) \transitiont{n+3} \cdots    
\end{align*}
of $\automaton'$, which starts in $(q_0, v_0)$, processes $ \rho \cdot_{t} \mu$, and is accepting.
Hence, $\languageof{\automaton'}$ is nonempty as required.

Conversely, let there be a $\mu \in \twty^\omega$ with $\rho \cdot_{t} \mu \in \languageof{\automaton'}$.
Then, there exists an accepting run 
\[(q_0, v_0) \transition{1} (q_1,v_1) \transition{2} (q_2,v_2) \transition{3} \cdots \]
of $\automaton'$ starting in some state~$(q_0, v_0)$ that processes $\rho \cdot_{t} \mu$, where $q_0 \in Q_0$ and with $v_0(x) =0$ for all $x$.
By definition of concatenation, there is an $n \ge 0$ with $t_n \le t$ and $t_{n+1} \ge t$ (where we use $t_0 = 0$ to allow $n = 0$) such that $\rho = (\sigma_1, t_1) \cdots (\sigma_1, t_n)$ and $\mu = (\sigma_{n+1}, t_{n+1}-t)(\sigma_{n+2}, t_{n+2}-t) \cdots$.
The run can be split into two parts:
\begin{itemize}
      \item The first part is
      \[(q_0, v_0) \transition{1} (q_1,v_1) \transition{2} \cdots \transition{n} (q_n,v_n), \]
      which witnesses 
      $(q_0, v_0) \xrightarrow{\rho}_{\automaton'} (q_n, v_n)$. Thus, $(q_n, v_n + (t - t_n)) \in \terminal{\automaton'}{t}$.
      
     \item The second part is 
     \begin{align*}
     & (q_n, v + (t - t_n)) \transitiontm{n+1} (q_{n+1}, v_{n+1}) \transitiontm{n+2} \\ &(q_{n+2}, v_{n+2}) \transitiontm{n+3}\cdots,
     \end{align*}
     which is an accepting infinite run of $\automaton'$ starting in $(q,v')$ that processes $\mu$.
\end{itemize}
Hence, $(q_n, v + (t - t_n)) \in \terminal{\automaton'}{\rho,t} \cap \nonempty{\automaton'}$, which is therefore, as required, nonempty.
\end{proof}

So far, this construction is very similar to the the classical procedure for monitoring \ac{ltl3} with the addition of $\tbadiv$ to account for time divergence.
However, the set of states with non-empty languages of \iac{tba} is likely to be infinite, and its reach-set over a symbolic trace (see Section~\ref{sec:uncertainty}) may be as well.
We now present a symbolic online algorithm to compute these infinite sets and their intersections in an efficient manner.

\subsection{Monitoring Algorithm}
We assume that the language~$\phi$ we will monitor and its complement $\bar{\phi}$ are given as \acp{tba} $\automaton_\phi$ and $\automaton_{\bar{\phi}}$, where $\languageof{\automaton_\phi} = \phi$ and $\languageof{\automaton_{\bar{\phi}}} = \bar{\phi}$.
We begin by computing the intersection of both automata with $\tbadiv$ (we hereafter refer to these intersection automata as $\automaton$ and $\compautomaton$).
We continue by finding the states of the automata with non-empty languages, also called the non-empty states from Definition \ref{def:non-empty}.
We then compute the intersection of the non-empty states with the reach-set (from Definition \ref{def:state-estimate}) of $\automaton$ and $\compautomaton$ over $\rho$.
If one of the intersections is empty, then we can output $\top$ or $\bot$, otherwise, we output $\unknown$.

We can calculate the set $\nonempty{\automaton}$ as a fixpoint using a backwards reachability algorithm. In order to practically work with the states of \iac{tba} we use a symbolic representation of the clock valuations, namely zones. A symbolic state $(q, Z)$ is a pair of a location~$q$ and a zone~$Z$.
A zone is a finite conjunction of clock constraints on the form $x_1 \sim n$ or $x_1 - x_2 \sim n$, where $x_1, x_2$ are clocks, $\sim \in \{<, \le, =, \ge, >\}$, and $n \in \nnrats$. 
Such a zone describes a convex set of clock valuations.

To compare observed time points with clock constraints in the zones, we add a global clock $x_G$ that will never be reset. 
Additionally we only consider rational values as inputs and in clock constraints, as we now discuss the implementation of our symbolic algorithm.
However, let us stress that a zone still includes all valuations satisfying the constraints, including real-valued clock valuations.
Zones may be efficiently represented using so-called \acp{dbm} \cite{DBLP:conf/ac/BengtssonY03}. 

We now define several zone operations that we will need.

\begin{definition} Given a zone $Z$ over a set of clocks $\clocks$ including the global clock time $x_G$, a set of clocks $\lambda \subseteq \clocks$, an interval $I = [t_1, t_2]$ between two positive rational numbers, and a clock constraint $g$, we define the following operations on zones: 
\begin{itemize}
    \item $\textit{free}_\lambda(Z) = \{v \mid \exists v' \in  Z \where \forall x \in \clocks \where v(x) = v'(x) \textit{ if } x \notin \lambda\}$
    \item $Z_\textit{free} = \textit{free}_\clocks(Z)$
    \item $Z[\lambda] = \{v \mid \exists v' \in Z \: \forall x \in \clocks \where  v(x) = 0 \textit{ if } x \in \lambda \textit{ otherwise } v(x) = v'(x) \}$
    \item $Z^\nearrow = \{v \mid \exists v' \in Z \where  v = v' + d  \textit{ for some } d \in \clockty\}$
    \item $Z^\swarrow = \{v \mid \exists v' \in Z \where  v = v' - d \textit{ for some } d \in \clockty\}$
    \item $Z^{\nearrow_I} = Z^\nearrow \land x_G \in I$
    \item $Z \land g = \{v \mid v \in Z \textit{ and } v \models g\}$
    \item $Z_0 = \{v \mid \forall x \in \clocks\where v(x) = 0\}$
\end{itemize}
Given a set~$S$ of symbolic states, $S^{\nearrow_{I}} = \{(q, Z^{\nearrow_{I}}) \mid (q, Z) \in S\}$ is obtained by applying the zone operation~$\nearrow_{I}$ to all zones in the set~$S$.
\end{definition}
All of the above operations on zones can be efficiently implemented using the \ac{dbm} data-structure \cite{DBLP:conf/ac/BengtssonY03}.

We now proceed to develop the online zone-based procedure we use to monitor real-time properties specified by \acp{tba}.
$Pred_\automaton(q, Z)$ (computed by Algorithm \ref{alg:pred}) is the set of symbolic states that can, by a single transition and delay, reach the state $(q, Z)$ of $\automaton$, i.e.,
\[
Pred_\automaton(q, Z) = \{(q', Z') \mid (q', q, \alpha, \lambda, g) \in \Delta \text{ and } Z' = \textit{free}_\lambda(Z^\swarrow \land \bigwedge_{x \in \lambda} x = 0 ) \land g\}
\]
\begin{algorithm}[ht]
\caption{Find the predecessors (single transition) of a state}
\label{alg:pred}
\begin{algorithmic}
 \Require{\iac{tba} $\automaton = (Q, Q_0, \Sigma, \clocks, \Delta, \mathcal{F})$ and a symbolic state $(q, Z)$}
 \Ensure{$\textit{Pred}_\automaton(q, Z)$}
 \State{$\textit{Predecessors} \gets \varnothing$}
 \For{$(q',q,\alpha,\lambda,g) \in \Delta$}
     \State{$Z' = \textit{free}_\lambda(Z^\swarrow \wedge \bigwedge_{x \in \lambda} x = 0) \wedge g$}
     \State{$\textit{Predecessors} \gets \textit{Predecessors} \cup \{(q', Z')\}$}
 \EndFor
 \State {\textbf{return} $\textit{Predecessors}$}
\end{algorithmic}
\end{algorithm}

$\textit{Reach}_\automaton(S)$ (computed by Algorithm \ref{alg:reach} implementing a classical backwards analysis (see, e.g., \cite{bouyer08})) is the set
\[
\textit{Reach}_\automaton(S) =\{
(q,v) \mid (q,v) \xrightarrow{\rho} (q',v') \text{ s.t.} (q',v') \in S \text{ and } \rho
 \text{ is nonempty}\}
\]
of symbolic states that can, by at least one transition, reach a state in $S$. 

\begin{algorithm}[ht]
\caption{Compute the states that can reach the given states with at least one transition}
\label{alg:reach}
\begin{algorithmic}
\Require{\iac{tba} $\automaton$ and a set of symbolic states $S$}
\Ensure{$\textit{Reach}_\automaton(S)$}
\State{$\textit{Waiting} \gets \varnothing$}
\State{$\textit{Passed} \gets \varnothing$}
\For{$s \in S$}
 \State{$\textit{Waiting} \gets \textit{Waiting} \cup \textit{Pred}_\automaton(s)$}
\EndFor
\While{$\textit{Waiting} \neq \varnothing$}
 \State{$s \gets \textit{Pop(Waiting)}$}
 \If{$s \notin \textit{Passed}$}
  \State{$\textit{Waiting} \gets \textit{Waiting} \cup \textit{Pred}_\automaton(s)$}
  \State{$\textit{Passed} \gets \textit{Passed} \cup \{s\}$}
 \EndIf
\EndWhile
\State{\textbf{return} $Passed$}
\end{algorithmic}
\end{algorithm}

For a TBA~$\automaton$ with set of locations~$Q$ and $Q' \subseteq Q$, let $\textit{Reach}_\automaton^\infty(Q')$ denote the set of states of $\automaton$ from which an infinite run starts that visits locations from $Q'$ infinitely often.
Algorithm \ref{alg:fixpoint} computes $\textit{Reach}_\automaton^\infty(Q')$ which is the limit set~$S_n = S_{n-1}$ where
\[
S_0 = \{(q, Z_\textit{free}) \mid q \in Q'\}
\]
is the set of symbolic states whose location is in $Q'$, and where 
\[
S_i = \textit{Reach}_\automaton(S_{i-1} \cap S_0)
\]
for $i>0$ is the set of states that can reach $S_{i-1} \cap S_0$ with at least one transition. 
Thus, from states in the limit~$S_n$, there is an infinite run that infinitely many times reaches a location in $Q'$. 

\begin{algorithm}[ht]
\caption{Calculate the set of states that can infinitely often reach a location in $Q'$}
\label{alg:fixpoint}
\begin{algorithmic}

\Require{\iac{tba} $\automaton$ and a set~ $Q'$ of locations of $\automaton$}
\Ensure{$\textit{Reach}_\automaton^\infty(Q')$}
\State{$S_{Q'} \gets \varnothing$}
\For{$q \in Q'$}
    \State{$S_{Q'} \gets S_{Q'} \cup \{(q, Z_\textit{free})\}$}
\EndFor
\State{$S_a \gets S_{Q'}$}
\State{$S_b \gets \varnothing$}
\While{$S_a \neq S_b$}
    \State{$S_b \gets S_a$}
    \State{$S_a \gets \textit{Reach}_\automaton(S_a \cap S_{Q'})$}
\EndWhile
\State{\textbf{return} $S_a$}

\end{algorithmic}
\end{algorithm}

We can use this to calculate the set of states, from which there is an accepting run:
Given \iac{tba}~$\automaton$, we write $\textit{Reach}_\automaton^\infty$ as a shorthand for $\textit{Reach}_\automaton^\infty(\mathcal{F})$, where $\mathcal{F}$ is the set of accepting locations of $\automaton$.

\begin{theorem}
\label{thm:sne}
Given \iac{tba} $\automaton$. Then $\textit{Reach}_\automaton^\infty = \nonempty{\automaton}$.
\end{theorem}

\begin{proof}
    Consider \iac{tba} with accepting states~$S_\mathcal{F}$ (i.e., those states with accepting locations) and let $\textit{AccPred}_\automaton(S) = \textit{Reach}_\automaton(S) \cap S_\mathcal{F}$ be the function mapping a set~$S$ of states to its accepting predecessors, i.e., to the set of accepting states that can, by one \emph{or} more transitions, reach a state in $S$.

    Notice that for any set~ $S$ of states we have $\textit{AccPred}_\automaton(S) \subseteq S$.
    The iterative process given by $S^0 = S_\mathcal{F}$ and $S^{n} = \textit{AccPred}_\automaton(S^{n-1})$ represents part of the transformation in the final loop of Algorithm~\ref{alg:fixpoint} if \(\mathcal{F}\) is given as the parameter. It will converge after a finite number of steps to a greatest fixed point $S^*$.
    This holds when using zones for the continuous part of the state space, as zones are finite unions of regions~\cite{alur1994tba}, which can only be reduced a finite number of times before becoming empty.
 
    The set $S^*$ contains all the states in $S_\mathcal{F}$ that can reach a state in $S^*$ by one or more transitions. 
    Thus, we have that $S^*$ is the set of accepting states that can infinitely often visit an accepting state.
    
    Then, $\textit{Reach}_\automaton(S^*)$ is the set of all states (not necessarily accepting) that can reach accepting states infinitely often, meaning $\textit{Reach}_\automaton(S^*) = \nonempty{\automaton}$.
\end{proof}

Using this fixpoint, we can implement online monitoring given $\automaton$ and $\compautomaton$ by storing the reach-set given by a finite timed word over $\automaton$ and $\compautomaton$, while continuously checking if the reach-sets still overlap with $\nonempty{\automaton}$ and $\nonempty{\compautomaton}$ respectively. If both reach-sets still have non-empty languages, then the verdict is $\unknown$, but if all the states in the reach-set of $\automaton$ (of $\compautomaton$) have empty languages, then the verdict is $\bot$ ($\top$, respectively).

\begin{algorithm}
\caption{Get the set of possible successor states from $S$ after an input $(\alpha, t)$}
\label{alg:successor}
\begin{algorithmic}

\Require{\iac{tba} $\automaton= ( Q, Q_0, \Sigma, \clocks, \Delta, \mathcal{F})$, a set of symbolic states $S$, and a timed input~$(\sigma, \tau) \in \Sigma \times \nnrats$}
\Ensure{$\textit{Succ}_\automaton^{S}(\sigma, \tau)$}
\State{$\textit{Successors} \gets \varnothing$}
\For{$(q, Z) \in S$}
    \For{$(q,q',\sigma,\lambda,g) \in \Delta$}
        \If{$Z^{\nearrow_{[\tau,\tau]}} \models g$}
            \State{$\textit{Successors} \gets \textit{Successors} \cup \{(q', (Z^{\nearrow_{[\tau,\tau]}} \land g)[\lambda])\}$}
        \EndIf
    \EndFor
\EndFor
\State{\textbf{return} \textit{Successors}}
\end{algorithmic}
\end{algorithm}

Given the procedure $\textit{Succ}_\automaton$ described in Algorithm \ref{alg:successor} we can compute the reach-set of $\automaton$ over a finite rational-timed word $\rho = (\sigma_1,\tau_1), (\sigma_2,\tau_2), \ldots (\sigma_n, \tau_n) \in \twty^*$ delayed up to a time-point $t \in \nnrats$. If $S_0$ is the set of initial states and $S_i = \textit{Succ}_\automaton^{S_{i-1}}(\alpha_i, t_i - t_{i-1})$, then  
$\terminal{\automaton}{\rho, t} = \{(q, Z^{\nearrow_{[t-\tau(\rho),t-\tau(\rho)]}}) \mid (q, Z) \in S_n\}$ is the reach-set after observing $\rho$ and time passing up to $t$.

Termination of Algorithm~\ref{alg:pred} and Algorithm~\ref{alg:successor} is straight forward, since they only contain bounded loops.
Finally, termination of Algorithm~\ref{alg:reach} has been shown in the proof of Theorem~\ref{thm:sne}.

\begin{algorithm}[ht]
\caption{Online monitoring procedure given $\automaton_\phi$ and $\automaton_{\bar{\phi}}$. Gives a verdict $\top$, $\bot$ or $\unknown$ after each input}
\label{alg:monitor}
\begin{algorithmic}
\State{$\automaton \gets \automaton_\phi \tbacap \tbadiv$}
\State{$\compautomaton \gets \automaton_{\bar{\phi}} \tbacap \tbadiv$}
\State{$S \gets \{(q,Z_0) \mid q\textit{ is an initial location of } \automaton\}$}
\State{$\bar{S} \gets \{(q,Z_0) \mid q\textit{ is an initial location of } \compautomaton\}$}
\Loop
\State{Receive new input: $i \in (\Sigma \times \nnrats) \cup \nnrats$}
\If{$i = (\sigma, \tau) \in \Sigma \times \nnrats$}
\State{$S \gets \textit{Succ}_{\automaton}^S(\sigma, \tau)$} 
\State{$\bar{S} \gets \textit{Succ}_{\compautomaton}^{\bar{S}} (\sigma, \tau)$}
\ElsIf{$i = t \in \nnrats$}
\State{$S \gets S^{\nearrow_{[t,t]}}$}
\State{$\bar{S} \gets \bar{S}^{\nearrow_{[t,t]}}$}
\EndIf
\If{$S \cap \textit{Reach}_{\automaton}^\infty = \emptyset$}
    \State{\textbf{output} $\bot$}
\ElsIf{$\bar{S} \cap \textit{Reach}_{\compautomaton}^\infty = \emptyset$}
    \State{\textbf{output} $\top$}
\Else
    \State{\textbf{output} $\unknown$}
\EndIf
\EndLoop
\end{algorithmic}
\end{algorithm}

\begin{theorem}
    Algorithm~\ref{alg:monitor} computes $\evalfuncdiv$ in the following sense: 
    Given a non-empty sequence of inputs in $\in \Sigma \times \nnrats \cup \nnrats$ such that all time points in the sequence are non-decreasing, let $\rho$ be the projection of the input sequence to $\Sigma \times \nnrats$, which is a timed word.
    Algorithm~\ref{alg:monitor} returns
    \begin{itemize}
        \item $\evaluatediv{\languageof{\automaton}}{\rho,\tau(\rho)}$, if the input sequence ends with an input from $\Sigma \times \nnrats$, and
        \item $\evaluatediv{\languageof{\automaton}}{\rho,t}$, if the input
        sequence ends with $t \in \nnrats$.
        
    \end{itemize}
\end{theorem}

\begin{proof}
Let $\rho = (\sigma_1,\tau_1),\ldots,(\sigma_n,\tau_n)$. An induction shows that when Algorithm~\ref{alg:monitor} is given inputs~$(\sigma_1,\tau_1),\ldots,(\sigma_n,\tau_n)$ step-by-step, then $S$ is equal to the reach-set~$\terminal{\automaton}{\rho,\tau(\rho)}$ and $\bar{S}$ is equal to the reach-set~$\terminal{\compautomaton}{\rho,\tau(\rho)}$.
Thus, when then given the input~$t$, $S$ is equal to $\terminal{\automaton}{\rho,t}$ and $\bar{S}$ is equal to $\terminal{\compautomaton}{\rho,t}$.
Hence, Theorem~\ref{thm:sne} and Theorem~\ref{thm:correctness} yield the desired result. 
\end{proof}

An overview of the online monitoring procedure (see Algorithm~\ref{alg:monitor}) with $\automaton_\phi$ and $\automaton_{\bar{\phi}}$ is as follows.
First we define $\automaton$ and $\compautomaton$ as the intersection of each input automaton with $\tbadiv$.
We then use the backwards reachability algorithm to compute the set of states that have a non-empty language (in each \ac{tba}). While continuously receiving inputs, we compute the symbolic successor states from the initial states. After each input, we check if there is an overlap between the states with a non-empty language, and the reach-sets and output a verdict. The verdict is either $\top$ or $\bot$ when one of the reach-sets falls outside the set of states with a non-empty language and $\unknown$ otherwise.

\begin{figure}[ht]
\centering
\begin{tikzpicture}

\node at (0,0) {
\begin{tikzpicture}
\node at (-.5,3.5) {$\automaton$};

\begin{scope}
    \clip[rounded corners] (0,0) rectangle (4,4);   
    \fill[gray!30,draw=gray!60,ultra thick] (0,0) -- (0,3.5) arc[start angle=90,end angle=0,radius=3.5] -- cycle;
\end{scope}
    \draw[rounded corners,ultra thick] (0,0) rectangle (4,4);

    \node[draw,circle,ultra thick] (1) at (.5,2) {};
    \node[draw,circle,ultra thick,inner sep = 7.5] (2) at (1.8,3) {};
    \node[draw,circle,ultra thick,inner sep = 6] (3) at (3.2,1.5) {};
    \node[draw,circle,ultra thick,inner sep = 5] (4) at (1.5,.5) {};

\path[->,thick]
(1) edge node[above,xshift=-.35cm] {\footnotesize $(\sigma_1,\tau_1)$} (2)
(2) edge node[left,xshift=-.0cm] {\footnotesize $(\sigma_2,\tau_2)$} (3)
(3) edge node[above,xshift=-.35cm] {\footnotesize $(\sigma_3,\tau_3)$} (4);
    
\end{tikzpicture}
};


\node at (7,0) {
\begin{tikzpicture}
\node at (-.5,3.5) {$\compautomaton$};

\begin{scope}
   \clip[rounded corners] (0,0) rectangle (4,4);   
    \fill[gray!30,draw=gray!60,ultra thick] (0,4) -- (0,1) arc[start angle=-90,end angle=0,radius=3] -- cycle;
\end{scope}
    \draw[rounded corners,ultra thick] (0,0) rectangle (4,4);    

    \node[draw,circle,ultra thick] (1) at (.5,2) {};
    \node[draw,circle,ultra thick,inner sep = 4] (2) at (1.5,3.5) {};
    \node[draw,circle,ultra thick,inner sep = 8.5] (3) at (2.6,2) {};
    \node[draw,circle,ultra thick,inner sep = 8] (4) at (3.5,.5) {};

    \path[->,thick]
        (1) edge node[above,xshift=-.4cm] {\footnotesize $(\sigma_1,\tau_1)$} (2)
        (2) edge node[very near start,right,xshift=-.0cm] {\footnotesize $(\sigma_2,\tau_2)$} (3)
        (3) edge node[left,xshift=-.0cm] {\footnotesize $(\sigma_3,\tau_3)$} (4);

\end{tikzpicture}
};

\end{tikzpicture}

\caption{Evolution of the reach-sets (depicted as circles) in $\automaton$ (left) and $\compautomaton$ (right) after $0$, $1$, $2$, and $3$ observations. The grey areas are the states with  non-empty language, i.e., $\textit{Reach}_{\automaton}^\infty$ and $\textit{Reach}_{\compautomaton}^\infty$.}
    \label{fig:monitor-proc}
\end{figure}

Figure \ref{fig:monitor-proc} shows an example of the state spaces of \acp{tba}~$\automaton$ and $\compautomaton$, with their non-empty states marked as grey and the reach-sets~$\terminal{\automaton}{(\sigma_1, \tau_1),\ldots, (\sigma_{i}, \tau_{i})}$ and $\terminal{\compautomaton}{(\sigma_1, \tau_1),\ldots, (\sigma_{i}, \tau_{i})}$ depicted as black circles. As long as both~$\terminal{\automaton}{(\sigma_1, \tau_1),\ldots, (\sigma_{i}, \tau_{i})}$ and $\terminal{\compautomaton}{(\sigma_1, \tau_1),\ldots, (\sigma_{i}, \tau_{i})}$ contain a (grey) state with non-empty language, the verdict is $\unknown$. After three observations, $\terminal{\compautomaton}{(\sigma_1, \tau_1),\ldots, (\sigma_{3}, \tau_{3})}$ contains no states with non-empty language, i.e., the verdict is $\top$.

\begin{example} 
Consider the \timelogic formula $\varphi = G_{\ge 0}(a \rightarrow F_{\leq 30} b)$ from Example~\ref{ex:response}.
We are given the \ac{tba} $\automaton$ in Figure~\ref{fig:tba-response} and its complement $\compautomaton$, the \ac{tba} for $\overline{\varphi} = F_{\ge 0}(a \wedge G_{\leq 30} \neg b)$ shown in Figure~\ref{fig:tba-response-comp}.
As before, we draw the sink state $\overline{q_4}$ here for illustrative purposes.
If we were to monitor $\languageof{\varphi}$, the first step would be to take the intersections $\automaton \tbacap \tbadiv$ and $\compautomaton \tbacap \tbadiv$, but we skip this step here because of the size of the resulting automata.

\begin{figure}[ht]
  \centering
  \begin{tikzpicture}[shorten >=1pt,node distance=3cm,on grid,auto,thick] 
  \node[state,initial] (q1)  {$\overline{q_1}$}; 
  \node[state,accepting,right=of q1] (q2) {$\overline{q_2}$};
  \node (dummy) [right =of q2] {};
  \node[state,accepting] (q3) [above= 1cm of dummy] {$\overline{q_3}$};
  \node[state] (q4) [below= 1cm of dummy] {$\overline{q_4}$};

    \path[->] 
    (q1) edge [loop above] node {$a,b,c$} ()
         edge node[above,align=center] {$a$\\$y \colonequals 0 $} (q2)
    (q2) edge [loop above] node {$a,c$} ()
         edge [bend left=30] node[above,align=center] {$a,b,c$\\$y>30$} (q3)
         edge [bend right=30] node[below,align=center] {$b$\\ $y \le 30$} (q4)
    (q3) edge [loop right] node {$a,b,c$} ()
    (q4) edge [loop right] node {$a,b,c$} ();
  \end{tikzpicture}
  \caption{\Ac{tba} corresponding to $F_{\ge 0}(a \wedge G_{\leq 30} \neg b)$}
  \label{fig:tba-response-comp}
\end{figure}

First we compute the sets of non-empty language states $\nonempty{\bar\automaton}$ and $\nonempty{\automaton}$ as:

\begin{itemize}
  \item $\nonempty{\automaton} = \{(q_1, \true), (q_2, x \le 30)\}$.
  \item $\nonempty{\bar\automaton} = \{(\bar{q_1}, \true), (\bar{q_2}, \true), (\bar{q_3}, \true)\}$.
\end{itemize}

Now suppose the finite prefix $\rho = (b,10),(a,20)$ as seen in Example~\ref{ex:response}. 
We compute the reach-set at time point 25 for $\automaton$ and $\bar\automaton$  as:
\begin{itemize}
  \item $\terminal{\automaton}{\rho, 25} = \{\ (q_2,x = 5\wedge x_G = 25)\ \}$.
  \item $\terminal{\compautomaton}{\rho, 25} = \{\ (\overline{q_1},y = 25\wedge x_G = 25), (\overline{q_2},y = 5 \wedge x_G = 25 )\ \}$.
\end{itemize}
Note that although our algorithm uses a symbolic representation for the reach-sets, for a concrete input the clock constraints are equalities. 
Also note that, for $\compautomaton$, there are two possible states for $\rho$ since $\compautomaton$ is non-deterministic.

Both these reach-sets intersect with the non-empty language states, which means there exist infinite extensions satisfying both properties. However, this is not true if we use a time point greater than 50. For example, the reach-set $\terminal{\automaton}{\rho, 51} = \{\ (q_2,x = 31\land x_G = 51)\ \}$ has an empty intersection with $\nonempty{\automaton}$, thus the verdict would be $\bot$.
\end{example}

\section{Timing Uncertainty}
\label{sec:uncertainty}

So far, we have assumed that inputs have rational- rather than real-valued time-points, but in reality it can be real-valued time-points with arbitrary mathematical precision. 
Instead of approximating a real-valued time-point $\tau_i$ as a rational value, we can assume that the time-points are observed with a certain precision. More accurately, we assume that we observe some interval~$I_i$ containing $\tau_i$, e.g., $[\lfloor \tau_i \rfloor, \lceil\tau_i\rceil]$.
In related settings,  concrete timing information may be missing for many reasons, particularly when monitoring distributed systems~\cite{jahanian1994time,rushby1999systematic,pike06note}.
The problem we consider is also closely related to the robustness of \ac{ta}~\cite{dewulf2008robust} as well as work on monitoring over unreliable channels~\cite{kauffman2019monitorability,kauffman2021what}.

We  assume that during  monitoring, we observe a \emph{symbolic} timed word of the form~$\rho_s = (\sigma, \upsilon)$, where $\sigma$ is a  finite word over the symbol alphabet $\Sigma$ and $\upsilon$ is a sequence~$I_1,I_2,\ldots,I_n$ of time intervals of the same length as $\sigma$.
Each time interval $I_i$ is a closed interval~$[l_i,u_i]$ with integer end-points representing a lower and upper bound of the real-valued time-point at which the symbol~$\sigma_i$ occurred.
We require $l_i \le l_{i+1}$ for all $i<n$.
The set of finite symbolic timed words is denoted $\twsymty^*$.
Note that, as we are concerned with an algorithmic solution, we limit bounds in symbolic timed words to natural numbers (which is equivalent to supporting rationals with a fixed granularity).

We define the concatenation of symbolic timed words $\rho_1$ and $\rho_2$ at time point~$t$ where $\rho_1 = (\sigma_1^1, [l_1^1, u_1^1]), (\sigma_2^1, [l_2^1, u_2^1]), ..., (\sigma_n^1, [l_n^1, u_n^1])$ is a finite word with $l_n^1 \le t$, 
and $\rho_2 = (\sigma_1^2, [l_1^2, u_1^2]), (\sigma_2^2, [l_2^2, u_2^2]), ...$ is a finite or infinite word, as $\rho_1 \cdot_t \rho_2 = (\sigma_1, [l_1, u_1]), (\sigma_2, [l_2, u_2]), ...$ such that 
\[
\sigma_i =
\begin{cases}
  \sigma_i^1 & \text{if } i \le n,\\
  \sigma_{i-n}^2 & \text{else,}
\end{cases}
\quad\text{and}\quad[l_i, u_i] = 
\begin{cases}
  [l_i^1, u_i^1] & \text{if } i \le n,  \\
  [l_{i-n}^2 + t, u_{i-n}^2 + t] & \text{else.}
\end{cases}
\]

Semantically, a symbolic timed word $\rho_s=(\sigma,I_1,I_2,\ldots,I_n)$ encodes all timed words~$(\sigma,\tau)$ of length~$n$ where $\tau_i\in I_i$. In this case, we write $\rho\sqsubseteq\rho_s$. 

\begin{remark}
As we require the lower bounds~$l_i$ of a symbolic timed word to be non-decreasing, each symbolic timed word~$(\sigma,I_1,I_2,\ldots,I_n)$ encodes at least one concrete timed word, i.e., the word~$(\sigma_1, l_1),\ldots,(\sigma_n, l_n)$.
\end{remark}

Also, whenever $\rho_s=(\sigma,I_1,I_2,\ldots,I_n)$ and $\rho'_s=(\sigma,J_1,J_2,\ldots,J_n)$  are two symbolic timed words, we write $\rho_s\sqsubseteq\rho'_s$ if $I_i\subseteq J_i$ for all $i=1\ldots n$. 

\begin{example}
Consider then symbolic timed word~$\rho_s = (b,[1,2]), (a,[5,6]), (c,[7,8])$.
Then we have $\rho=(b,1.2), (a,5.4), (c,7.3) \sqsubseteq\rho_s$.
\end{example}

Monitoring a language of timed infinite words in the setting of timing uncertainty refines timed monitoring in the following way.  Given a finite symbolic prefix, it is checked whether all concrete realizations of this prefix determine the property.  That is whether all possible infinite extensions of such a concrete realization are included in the monitored property. 

\begin{definition}[Monitoring with timing uncertainty]
Given a language of infinite timed words $\lang \subseteq \twty^\omega$, a finite symbolic timed word $\rho_s \in \twsymty^*$ ending in $(\sigma,[l,u])$, and a $t \in \nnreals$ such that $u \le t$, the function $\evalfunc_U : \powerset{\twty^\omega} \rightarrow \twsymty^* \times \nnreals \rightarrow \boolthreety$ evaluates to a verdict with the following definition:
\[
\evalfunc_U(\lang)(\rho_s, t) = \left.
  \begin{cases}
    \top & \text{if $\rho \concat_t \mu \in \lang\ $ for all $ \rho\sqsubseteq\rho_s$ and all $ \mu \in \twdivty^{\omega}$}, \\
    \bot & \text{if $\rho \concat_t \mu \notin \lang\ $ for all $ \rho\sqsubseteq\rho_s $ and all  $ \mu \in \twdivty^{\omega}$}, \\
    \unknown & \text{otherwise}.
  \end{cases}
  \right.
\]
If $u > t$, then $\evalfunc_U(\lang)(\rho_s, t)$ is undefined.
\end{definition}

\begin{remark}
If $\rho_s\sqsubseteq\rho'_s$ then $\evalfunc_U(\lang)(\rho_s,t)\truthvaluegeq\evalfunc_U(\lang)(\rho'_s,t)$, where we define the partial order~$\truthvaluegeq$ over the set~$\set{\top, \unknown, \bot}$ such that $\top\truthvaluegeq\unknown$ and $\bot\truthvaluegeq\unknown$, but $\top$ and $\bot$ are incomparable. Intuitively, $\truthvaluegeq$ captures that  definitive verdicts are more informative than the inconclusive verdict~$\unknown$.
\end{remark}

\begin{example}
Consider the \timelogic property $F_{[5,6]} a$ ``an '$a$' has to occur within 5 to 6 time-units after the initial observation'', and the concrete timed word $\rho=(b,0), (b,1.2), (a,5.4), (c,7.3)$. Assume that we observe time-points as integer-bounded intervals of length one respectively length two (except for the initial observation) reflected by the following two symbolic timed words 
\[\rho_s^1 =(b,0), (b,[1,2]), (a,[5,6]), (c,[7,8])\] 
and 
\[\rho_s^2 =(b,0) (b,[1,3]), (a,[5,7]), \\(c,[7,9]).\]
Then, we have  
$\evalfuncdiv(\languageof{F_{[5,6]})a})(\rho,8)=\top$ 
as well as 
$\evalfunc_U(\languageof{F_{[5,6]})a})(\rho_s^1,8)=\top$, 
whereas 
$\evalfunc_U(\languageof{F_{[5,6]})a})(\rho_s^2,8)=\unknown$, as there are concrete timed words encoded by $\rho_s^2$ that have an $a$ in the interval~$[5,6]$ and there are concrete timed words encoded by $\rho_s^2$ that do not have an $a$ in the interval~$[5,6]$, e.g., if the $a$ occurs at time~$7$.
\end{example}

To obtain a monitoring algorithm for monitoring under timing uncertainty it merely suffices to extend the symbolic successor computation in Algorithm~\ref{alg:successor} to pairs~$(\alpha,I)$ where $I$ is an integer-bounded interval. Since we defined the delay operation for intervals the only necessary change is to replace $Z^{\nearrow_{[\tau, \tau]}}$ with $Z^{\nearrow_I}$ in the innermost \textit{if}-statement of Algorithm~\ref{alg:successor}. Thus, the monitoring procedure (Algorithm~\ref{alg:monitor}) can easily be extended to support timing uncertainty.

\begin{theorem}
    There is an online algorithm that computes $\evalfunc_U$.
\end{theorem}


\lstdefinestyle{console}{
    frame=single,
    basicstyle=\ttfamily\small,
    breakatwhitespace=true,
    breaklines=true,
    captionpos=b,
    keepspaces=true,
    numbers=left,
    numbersep=5pt,
    showspaces=false,
    showstringspaces=false,
    showtabs=false,
    tabsize=4,
}

\def\console{\lstinline[style=console]}

\section{Tool Implementation: \monitaal}

To implement the monitoring algorithms presented in Sections~\ref{sec:solution} and \ref{sec:uncertainty}, we developed the tool \monitaal\footnote{\url{https://github.com/DEIS-Tools/MoniTAal}}. 
\monitaal takes two \acp{tba} as input. Each \ac{tba} has to be the negation of the other. The user can then either monitor a log of observations, or include the monitor as a library in their own software, and monitor observations in an online fashion. A verdict ($\top$, $\bot$, or  $\unknown$) is output after each observation.

In this section we demonstrate and benchmark \monitaal. We monitor properties from a verified gear controller model developed in collaboration between Lindahl~\etal\cite{gear} and an automotive software company. In addition, we benchmark \monitaal combined with the formula translation tool \mitppl, on properties from \timescales.

\monitaal follows the online monitoring procedure described in Algorithm~\ref{alg:monitor} by first computing the non-empty language states of each input automaton and then in an online fashion, for each observation, updating the reach-set and giving a verdict in $\{\top, \bot, \unknown\}$. When a verdict $\top$ or $\bot$ is given, the monitoring procedure stops (since no further observations can change the verdict).  

\subsection{The DBM Library~\textsc{Pardibaal}}

Computing the non-empty language states, as well as the reach-sets from interval-timed traces, requires handling states symbolically using zones. To do this, we implemented the \ac{dbm} library \textsc{Pardibaal}\footnote{\url{https://github.com/DEIS-Tools/PARDIBAAL}} to handle operations on zones. The existing \ac{dbm} library UDBM\footnote{\url{https://github.com/UPPAALModelChecker/UDBM}} is very efficient and currently used by the model checker \textsc{Uppaal}, but complex to maintain and released under the GPLv3 license.
We implemented \textsc{Pardibaal} to have \iac{dbm} library under the LGPL licence, permitting inclusion in proprietary applications, as well as to have a modern implementation that is easier to maintain.

\textsc{Pardibaal} implements all standard \ac{dbm} operations~\cite{DBLP:conf/ac/BengtssonY03} and additional functionality such as subtraction, intersection, and specialized delay operations. In practice, we work with sets of symbolic states and therefore have multiple \acp{dbm} (zones). Most of the basic operations are trivially extended to finite unions of zones (i.e., finite sets of \acp{dbm}) called federations. Equality or inclusion checking is more complicated using federations, since a single zone could be a subset of a union of multiple zones, without being a subset of any one of them on their own. An exact implementation requires an expensive difference/subtraction operation between \acp{dbm}. One way to approximate the relation between federations is to combine the relation between all pairs of \acp{dbm}.
The subtraction operation is also important in the backwards analysis part used to compute the set of non-empty language states.

\subsection{Demonstration of \monitaal}

Consider the property
\[ \varphi_{a} = F_{\le 10}(a) \land G_{\ge 0}(a\rightarrow F_{]0,10]}(a)). \]
$\varphi_{a}$ requires $a$ to be observed infinitely often, with at least one $a$ being observed within 10 time-units of the first observation, and then the time between two consecutive observations of $a$ is no more than 10. 
Note that when $a$ has not been observed for more than ten units of time, then the verdict~$\bot$ can be given. On the other hand, the verdict~$\top$ will never be given when monitoring~$\varphi_a$.

Now, consider the property 
\[\varphi_{b} = F_{\ge 0}(G_{\ge 0}(\neg b)).\] 
$\varphi_b$ requires that ``from some point onwards, we will never see $b$ again''. 
Note that monitoring $\varphi_{b}$ will never result in a $\top$ or $\bot$ verdict, since there is always an infinite extension satisfying the property and another one that does not.

In the following example we monitor the  property:
\[\varphi_{ab} = \varphi_{a} \land \varphi_{b}.\]
When monitoring the property~$\varphi_{ab}$ one again obtains the verdict~$\bot$ when $a$ has not been observed for more than ten units of time, but the verdict~$\top$ is never given.
The \acp{tba} accepting the languages of $\varphi_{ab}$ and $\neg\varphi_{ab}$ are shown in \autoref{fig:bench-aut}.
\begin{figure}[ht]
  \centering
\begin{tikzpicture}[shorten >=1pt,node distance=2cm,on grid,auto, thick] 
  \node[state] (q1)  {$p_2$}; 
  \node[state,accepting] (q2) [right=3cm of q1] {$p_3$};
  \node[state,initial] (q0) [below=3cm of q1] {$p_1$};
  
    \path[->]
    (q0) edge node[align=center, right] {$a,b$\\$x \colonequals 0$} (q1)
    (q1) edge [loop left] node {$b$} ()
    (q1) edge [loop above] node[align=center] {$a$\\$x\le 10$\\$x\colonequals 0$} ()
         edge node[above,align=center] {$a$\\$x\le 10$\\$x \colonequals 0$} (q2)
    (q2) edge [loop above] node[align=center] {$a$\\$x\le 10$\\$x \colonequals 0$} ();

\node[state] (q3) [right=3cm of q2]  {$n_2$}; 
\node[state,accepting] (q4) [right=4cm of q3] {$n_3$};
\node (dummy) [right=2cm of q3] {};
\node[state,accepting] (q5) [below=3cm of q4] {$n_4$};
\node[state,initial] (q01) [below=3cm of q3] {$n_1$};

    \path[->]
    (q01) edge node[left,align=center] {$a,b$\\$x \colonequals 0$} (q3)
    (q3) edge [loop above] node[align=center] {$a$\\$x\le 10$\\$x\colonequals 0$} ()
    (q3) edge [bend right=25] node[below,align=center] {$b$} (q4)
    (q3) edge node[below,align=center] {$a$\\$x> 10$} (q5)
    (q4) edge node[above,align=center] {$b$} (q3)
    (q4) edge [bend right=25] node[above,align=center] {$a$\\$x\le 10$\\$x\colonequals 0$} (q3)
    (q4) edge node[right,align=center] {$a$\\$x> 10$} (q5)
    (q5) edge [loop below] node[below] {$a,b$} ()
    ;
\end{tikzpicture}
  
  \caption{\Acp{tba} accepting the languages of $\varphi_{ab}$ (left) and $\neg\varphi_{ab}$ (right).}
  \label{fig:bench-aut}
\end{figure}

During monitoring, the reach-sets are represented as lists~$[(q_1, Z_1), (q_2, Z_2), \ldots, (q_k, Z_k)]$ of symbolic states where, for all $i \in [1, k]$, $q_i$ is a location and $Z_i$ is a zone.
In our experiments, we want to evaluate the size of the representation of the reach-sets (corresponding to the $k$).
When monitoring a concrete trace (a trace where observation times are singular intervals), each $Z_i$ in the symbolic state  encodes a single clock valuation. In addition, if the monitored \ac{tba} is deterministic then $k$ will always be $1$. 
Note that in the case of non-determinism and/or uncertain traces (traces where observation times are intervals) the size of the representation of the reach-set can increase.
As previously mentioned, we represent zones as \acp{dbm}, but when monitoring concrete traces we represent the clock valuation as a list of assignments rather than \iac{dbm}.

We say that the reach-set representation~$[(q_1, Z_1), (q_2, Z_2), \ldots, (q_k, Z_k)]$ is reduced w.r.t.\ zone-inclusion, if whenever $q_i=q_j$ for $i\ne j$, then $Z_i \not\subseteq Z_j$ and $Z_j \not\subseteq Z_i$.

\begin{example}
  Let us monitor the automata from \autoref{fig:bench-aut} over a concrete trace $\rho = (\sigma_1, \tau_1), (\sigma_2, \tau_2),\ldots,(\sigma_n, \tau_n)$ containing a series of $a$'s and $b$'s using the reduced state space representation. Looking at just the automaton for $\varphi_{ab}$, the reach-set~$\terminal{\automaton_{\varphi_{ab}}}{\rho, \tau_n}$ has one of the following three forms:
  
  \begin{enumerate}
    \item $\emptyset$ if the distance between two $a$'s is greater than 10, or
    \item $\{(p_0, \{x = 0, x_G = \tau_n\}), (p_1, \{x = 0, x_G = \tau_n\})\}$ if $\sigma_n = a$, or
    \item $\{(p_0, \{x = \tau_n - \tau_{n-1}, x_G = \tau_n\})\}$ if $\sigma_n = b$.
  \end{enumerate}

\end{example}

Now we will demonstrate \monitaal monitoring $\varphi_{ab}$ over short traces. \autoref{code:concrete} shows an execution of the trace $(a, 0), (a, 4), (b, 20)$.
The trace is entered incrementally on lines 9, 17, and 26. The outputs between the inputs are the reach-sets (as long as the verdict is $\unknown$) or a definitive verdict is given, at which point the algorithm terminates. 
The ``positive'' and ``negative'' reach-sets refer to the reach-set of the \ac{tba} accepting the language of the monitored property and its negation. The representation of the reach-set is printed as a location with a valuation for all clocks where \console{0} is the ``zero clock'' and \console{global} is the global clock previously referred to as $x_G$.

We see that the reach-sets of the two automata together contain three states after observing $(a, 4)$ on line 17, which is the maximum size of these sets in this execution.
After observing $(b, 20)$ on line 26 the property can no longer be satisfied, so the monitoring ends with a $\bot$ verdict printed as \console{NEGATIVE}.

\begin{figure}[ht]
    \centering
\begin{lstlisting}[style=console]
$ ./monitaal-bin <...>

Positive reach-set:
p1 : 0 = 0, x = 0, global = 0

Negative reach-set:
n1 : 0 = 0, x = 0, global = 0

Next event: @0 a

Positive reach-set:
p2 : 0 = 0, x = 0, global = 0

Negative reach-set:
n2 : 0 = 0, x = 0, global = 0

Next event: @4 a

Positive reach-set:
p2 : 0 = 0, x = 0, global = 4
p3 : 0 = 0, x = 0, global = 4

Negative reach-set:
n2 : 0 = 0, x = 0, global = 4

Next event: @20 b
Monitoring ended, verdict is: NEGATIVE
Monitored 3 events
\end{lstlisting}
\caption{Example of monitoring the property~$\varphi_{ab}$ over the concrete timed trace $(a, 0), (a, 4), (b, 20)$.}
\label{code:concrete}
\end{figure}

\autoref{code:interval} is a similar example, except that the trace has uncertain timing. The trace monitored here is $(a,0), (a, \lbrack0, 10\rbrack), (a, \lbrack1, 11\rbrack), (b, \lbrack30, 30\rbrack)$. The clock values are now handled symbolically as zones, instead of just as single valuations. A zone here is printed as \iac{dbm}, where rows and columns refer to clocks. The first row and column is the ``zero clock'' (the value of which is always zero), the second is the clock $x$ from the automaton, and the third is the global clock (\console{global}/$x_G$). An entry $(\prec, n)$ on row $i$ and column $j$ of \iac{dbm} is interpreted as the constraint $x_i - x_j \prec n$ where $\prec \in \{<, \le\}$ and $n$ is an integer. For example, on line 29 one can see that the two first entries of that row indicate the following constraints: $x_G - 0 \le 10$ and $x_G - x \le 10$.

Notice that after observing $(a, [20, 30])$ (line 52) the negative reach-set includes two symbolic states in $n_4$, where the difference between them are clock constraints on $x$. It is actually unnecessary to store clock constraints on $x$ in this location, since the value of $x$ will at no point in the future be checked by a guard.
Because of this, we could ignore the value of this clock, so that all states in $n_4$ become equivalent. 

This is an application of the inactive clock-abstraction by Daws and Yovine~\cite{DBLP:conf/rtss/DawsY96}. By analysing the automaton, we can find the set of active clocks for each location. A clock is active if there is a path where the clock appears in a guard before being reset. If a clock is not active, it can be ignored. This allows us to prune some symbolic states that, even though they have different clock constraints, will not impact the result. Note that we can never ignore the zero or global clock, because they are actively used in the procedure.

\begin{figure}[ht]
\centering
\begin{tikzpicture}[every node/.style={inner sep=4,outer sep=0}]
\node at (0,0) {\begin{minipage}{.4\linewidth}\begin{lstlisting}[style=console, frame={tlr}, basicstyle=\ttfamily\footnotesize]
$ ./monitaal-bin <...>
...

Next event: @[0,0] a

Positive reach-set:
p2 
<<<<<<
<=0     <=0     <=0     
<=0     <=0     <=0     
<=0     <=0     <=0     
>>>>>>

Negative reach-set:
n2 
<<<<<<
<=0     <=0     <=0     
<=0     <=0     <=0     
<=0     <=0     <=0     
>>>>>>

Next event: @[0, 10] a

Positive reach-set:
p2 
<<<<<<
<=0     <=0     <=0     
<=0     <=0     <=0     
<=10    <=10    <=0     
>>>>>>
p3 
<<<<<<
<=0     <=0     <=0     
<=0     <=0     <=0     
<=10    <=10    <=0     
>>>>>>

Negative reach-set:
n4 
<<<<<<
<=0     <-10    <-10    
<=15    <=0     <=0     
<=15    <=0     <=0     
>>>>>>
n2 
\end{lstlisting}\end{minipage}\hspace{7mm}\begin{minipage}{.6\linewidth}\begin{lstlisting}[firstnumber=46, style=console, frame={blr}, basicstyle=\ttfamily\footnotesize]
<<<<<<
<=0     <=0     <=0     
<=0     <=0     <=0     
<=10    <=10    <=0     
>>>>>>

Next event: @[20, 30] a

Positive reach-set:
p2 
<<<<<<
<=0     <=0     <=-20   
<=0     <=0     <=-20   
<=20    <=20    <=0     
>>>>>>
p3 
<<<<<<
<=0     <=0     <=-20   
<=0     <=0     <=-20   
<=20    <=20    <=0     
>>>>>>

Negative reach-set:
n4 
<<<<<<
<=0     <=-20   <=-20   
<=30    <=0     <=0     
<=30    <=0     <=0     
>>>>>>
n4 
<<<<<<
<=0     <-10    <=-20   
<=30    <=0     <=0     
<=30    <=10    <=0     
>>>>>>
n2 
<<<<<<
<=0     <=0     <=-20   
<=0     <=0     <=-20   
<=20    <=20    <=0     
>>>>>>

Next event: @[41, 45] b   
Monitoring ended, verdict is: NEGATIVE
Monitored 4 events
\end{lstlisting}
\end{minipage}
};
\end{tikzpicture}
\caption{Monitoring the property~$\varphi_{ab}$ over the uncertain timed trace $(a, 0), (a, \lbrack0,15\rbrack), (a, \lbrack20,30\rbrack), (b, \lbrack41,45\rbrack)$.}
\label{code:interval}
\end{figure}

Following the example, it is obvious that the complexity highly depends on the monitored property (and how concise the property automaton is constructed) as well as the observations. For the simple response property seen in \autoref{ex:response}, the time and memory usage per observation is negligible (below 100 microseconds). 
This is mainly because the property automata are small, and the inactive clock-abstraction implies an upper bound on the size of the representation of the reach-sets in those cases. Since we are doing online monitoring, the previous observations do not take up space in memory.
The response time between verdicts can increase more when the representation of the reach-set becomes large. We say that the response time is the time it takes between providing an observation to receiving a verdict.

To demonstrate monitoring properties and providing verdicts in a more realistic scenario we will examine the gear controller model by \cite{gear}.

The gear controller model is separated into several components, one of which is the interface. The interface sends requests \textit{ReqNewGear} after which the gear controller changes the gear and sends a response \textit{NewGear}.
The expected time it takes to change gear depends on whether or not the gear is being changed to or from neutral, and if some non-critical failures are observed in the engine.
To describe this we take the conjunction of two properties that distinguish between gear-change requests between neutral and non-neutral gears.
The request $\textit{ReqNewGear}_\textit{neu}$ is used when the gear change is to or from neutral, while $\textit{ReqNewGear}_k$ is used otherwise. $\textit{error}_1$ and $\textit{error}_2$ refer to observations in the engine signaling problems with matching torque and speed.

\newpage
\begin{align*}
\varphi_{\textit{gear-neu}}
= & \ G_{\ge 0} (\textit{ReqNewGear}_{neu} \rightarrow (X_{[150, 900]}(NewGear))\\ 
& \lor (X_{[150, 900]} (\textit{error}_1) \land F_{[550, 1055[}(\textit{NewGear}))\\
& \lor (X_{[150, 900]} (\textit{error}_2) \land F_{[450, 1205]}(\textit{NewGear})))
\end{align*}

\begin{align*}
\varphi_{\textit{gear-k}} 
= & \ G_{\ge 0} (\textit{ReqNewGear}_k \rightarrow (X_{[400, 900]}(NewGear))\\ 
& \lor (X_{[150, 900]} (\textit{error}_1) \land F_{[700, 1055[}(\textit{NewGear}))\\
& \lor (X_{[150, 900]} (\textit{error}_2) \land F_{[750, 1205]}(\textit{NewGear})))
\end{align*}

The combined property $\varphi_\textit{gear} = \varphi_\textit{gear-neu} \land \varphi_\textit{gear-k}$ describes the timing guarantees of the gear controller when changing gears.
To monitor the property, we simulate the behavior of the gear controller. Starting in neutral, a request is sent to change the gear up or down. A response is observed at some time point, depending on whether or not a non-critical error occurred.

For this demonstration we simulate a gear controller that can change gears 150 time units earlier or later than the specification allows. The simulation selects the timing of the gear change uniformly. Other than the timing, the gear controller behaves correctly. Furthermore, the behavior is observed with a timing uncertainty, meaning that all timing information of observations are intervals of 100 time units. Note that the simulation of the faulty gear controller is not guaranteed to exhibit a faulty gear change at any point.

Fig.~\ref{fig:histogram} shows a histogram reporting how many times a gear change was observed, before an error was detected, across 1000 simulations of the faulty gear controller. As expected, the chance of no error occurring decreases. The verdict given in all these cases are simply $\bot$, since satisfaction of the property $\varphi_\textit{gear}$ cannot be determined by a finite prefix. The demonstration shows that in all 1000 cases an error was detected. The longest trace showed an error after 65 gear changes. 

\begin{figure}[ht]
    \centering
    \begin{tikzpicture}
    \begin{axis}[
        height=0.5\linewidth,
        width=\linewidth,
    	x tick label style={
    		/pgf/number format/1000 sep=10},
    	ylabel=Observations,
    	enlargelimits=0.03,
    	ybar=0pt,
        bar width=1,
    ]
    \addplot coordinates {(1,98) (2,94) (3,85) (4,89) (5,64) (6,70) (7,60) (8,70) (9,38) (10,57) (11,28) (12,29) (13,20) (14,20) (15,14) (16,24) (17,23) (18,9) (19,15) (20,9) (21,11) (22,7) (23,7) (24,6) (25,9) (26,5) (27,6) (28,6) (29,3) (30,5) (31,2) (32,2) (33,2) (34,0) (35,1) (36,2) (37,3) (38,2) (39,0) (40,0) (41,0) (42,1) (43,0) (44,0) (45,0) (46,0) (47,1) (48,1) (49,0) (50,0) (51,1) (52,0) (53,0) (54,0) (55,0) (56,0) (57,0) (58,0) (59,0) (60,0) (61,0) (62,0) (63,0) (64,0) (65,1)};
    \end{axis}
    \end{tikzpicture}
    \caption{Histogram of the number of gear changes observed before a fault was detected over 1000 runs.}
    \label{fig:histogram}
\end{figure}
\subsection{Performance}
We run experiments of \monitaal on the property~$\varphi_{ab}$ described above, as well as the gear controller property $\varphi_{gear}$ taken from \cite{gear}.
For these experiments the properties $\varphi_\textit{gear}$ and $\varphi_\textit{ab}$ are monitored over traces of three different lengths, and are monitored with combinations of timing uncertainty and under time divergence.

\autoref{results2} shows the experimental results for monitoring the property~$\varphi_{ab}$. 
The results are shown for two different traces of (parametric) length~$k$:
\begin{enumerate}
  \item $\rho_c = (a, 1), (a, 2), \ldots, (a, k-1), (b, 30+k)$
  \item $\rho_u = (a [0, 10]), (a, [1, 11]), \ldots, (a, [k-1, k + 9]), (b, [k+30, k+30])$
\end{enumerate}
$\rho_c$ is a concrete trace while $\rho_u$ is an uncertain trace.

The results show us that we, in all cases, obtain a bound on the size of the reach-set. This is very useful, since the size of the reach-set is what primarily determines the response time. 
However, the maximum number of \emph{different} symbolic states is determined by the construction of the 
\acp{tba}. This can be observed by looking at the impact of monitoring under time divergence in the case of uncertain timing. When monitoring under time divergence we use \ac{tba}-intersection, resulting in much larger \acp{tba}, thus increasing the maximum  number of possible symbolic states. 

The observed trace can also affect the size of the reach-set, because observations can enable a different number of transitions. This can either be because of non-determinism, or because of uncertain timing as seen in the difference between the result of $\rho_c$ and $\rho_u$.

  \begin{table}[ht]
    \begin{tabular}{rrrr}
      \toprule
      \textbf{Word length} & \textbf{Total time} & \textbf{Max response} & \textbf{Max Size} \\\midrule
      \multicolumn{4}{l}{\textbf{$\rho_c$}}\\

    25,000 &   43.0 $\pm$ 0.6 ms & 25.9 $\pm$ 8.1 $\mu$s & 3\\
    50,000 &   86.0 $\pm$ 0.9 ms & 31.1 $\pm$ 9.0 $\mu$s & 3\\
    100,000 &   172.3 $\pm$ 3.1 ms & 33.2 $\pm$ 7.7 $\mu$s & 3\\
    200,000 &   343.4 $\pm$ 2.0 ms & 38.2 $\pm$ 5.2 $\mu$s & 3\\\midrule
    \multicolumn{4}{l}{\textbf{$\rho_c$ w. time divergence}}\\
    25,000 &   105.8 $\pm$ 0.9 ms & 25.2 $\pm$ 6.6 $\mu$s & 4\\
    50,000 &   211.8 $\pm$ 2.8 ms & 30.3 $\pm$ 11.2 $\mu$s & 4\\
    100,000 &   423.1 $\pm$ 4.0 ms & 30.4 $\pm$ 8.4 $\mu$s & 4\\
    200,000 &   849.1 $\pm$ 11.2 ms & 33.5 $\pm$ 7.0 $\mu$s & 4\\\midrule
    \multicolumn{4}{l}{\textbf{$\rho_{u}$}}\\
    25,000 &   150.9 $\pm$ 0.6 ms & 31.3 $\pm$ 7.8 $\mu$s & 3\\
    50,000 &   301.8 $\pm$ 1.9 ms & 30.3 $\pm$ 7.2 $\mu$s & 3\\
    100,000 &   605.0 $\pm$ 5.9 ms & 33.9 $\pm$ 5.4 $\mu$s & 3\\
    200,000 &   1,208.0 $\pm$ 4.3 ms & 36.3 $\pm$ 7.8 $\mu$s & 3\\\midrule
    \multicolumn{4}{l}{\textbf{$\rho_{u}$ w. time divergence}}\\
    25,000 &   2,959.2 $\pm$ 16.9 ms & 166.8 $\pm$ 6.4 $\mu$s & 16\\
    50,000 &   5,926.7 $\pm$ 40.6 ms & 170.6 $\pm$ 8.0 $\mu$s & 16\\
    100,000 &   11,861.5 $\pm$ 47.0 ms & 170.8 $\pm$ 5.5 $\mu$s & 16\\
    200,000 &   23,760.1 $\pm$ 123.8 ms & 173.2 $\pm$ 8.1 $\mu$s & 16\\\bottomrule
    
    \end{tabular}
    \caption{Results for monitoring $\varphi_{ab}$. Here, \myquot{Total time} refers to to the overall runtime of the algorithm, \myquot{Max response} refers to the maximal time between a new observation and the corresponding verdict, and \myquot{Max Size} refers to the maximal size of the reach-sets of both automata together. The results are the mean values of 50 runs.}
    \label{results2}
  \end{table}

\autoref{results_gear} shows the results of monitoring the gear controller properties in scenarios similar to the $\varphi_{ab}$ experiment.
In this case the maximum reach-set size does not change when monitoring a concrete trace under time divergence. This is because both the property automata and the time divergence automaton are deterministic. Thus under intersection, the automaton is still deterministic. Monitoring a deterministic automaton with precise/concrete observations will always result in a singleton reach-set.
However timing uncertainty does increase the size of the reach-set, and even more so when also combined with time divergence.


  \begin{table}[ht]
    \begin{tabular}{rrrr}
      \toprule
      \textbf{Word length} & \textbf{Total time} & \textbf{Max response} & \textbf{Max Size} \\\midrule
      
        
      
      


      \multicolumn{4}{l}{\textbf{Gear Controller}}\\
    1000 & 4.1 $\pm$ 0.0 ms  & 14.9 $\pm$ 9.3 $\mu$s & 2\\
    5000 & 20.4 $\pm$ 0.5 ms  & 21.4 $\pm$ 9.2 $\mu$s & 2\\
    10000 & 40.4 $\pm$ 0.5 ms  & 20.7 $\pm$ 5.3 $\mu$s & 2\\\midrule
    \multicolumn{4}{l}{\textbf{Gear Controller w. time divergence}}\\
    1000 & 9.8 $\pm$ 0.1 ms  & 25.3 $\pm$ 7.6 $\mu$s & 2\\
    5000 & 49.0 $\pm$ 0.8 ms  & 28.8 $\pm$ 6.7 $\mu$s & 2\\
    10000 & 97.6 $\pm$ 0.6 ms  & 30.3 $\pm$ 6.6 $\mu$s & 2\\\midrule
    \multicolumn{4}{l}{\textbf{Gear Controller w. uncertainty}}\\
    1000 & 6.8 $\pm$ 0.1 ms  & 18.8 $\pm$ 6.7 $\mu$s & 3\\
    5000 & 34.0 $\pm$ 0.3 ms  & 22.6 $\pm$ 4.7 $\mu$s & 3\\
    10000 & 68.1 $\pm$ 0.6 ms  & 22.5 $\pm$ 5.7 $\mu$s & 3\\\midrule
    \multicolumn{4}{l}{\textbf{Gear Controller w. uncertainty and time divergence}}\\
    1000 & 41.2 $\pm$ 0.4 ms  & 81.1 $\pm$ 2.9 $\mu$s & 12\\
    5000 & 209.5 $\pm$ 1.9 ms  & 87.5 $\pm$ 6.1 $\mu$s & 13\\
    10000 & 421.4 $\pm$ 4.9 ms  & 88.2 $\pm$ 5.4 $\mu$s & 13\\\bottomrule

    \end{tabular}
    \caption{Results for monitoring the $\varphi_{\textit{gear}}$ over a simulation of the gear controller model from \cite{gear}. Here, \myquot{Total time} refers to to the overall runtime of the algorithm, \myquot{Max response} refers to the maximal time between a new observation and the corresponding verdict, and \myquot{Max Size} refers to the maximal size of the reach-sets of both automata together. The results are the mean values of 50 runs.}
    \label{results_gear}
  \end{table}

\subsection{Time Divergence}
Running the tool on the property~$F_{\ge 20}a$ from \autoref{ex:gt20}  illustrates how intersection with the time divergence automaton can give an immediate verdict. \autoref{code:past20} shows \monitaal being run with and without time divergence (using option \console{--div <alphabet>}). As expected, a verdict is provided instantly only when time divergence is taken into account.

\begin{figure}[ht]
    \centering
\begin{lstlisting}[style=console]
$ ./MoniTAal-bin <...>

Positive reach-set:
q1 : 0 = 0, x = 0, global = 0

Negative reach-set:
q1 : 0 = 0, x = 0, global = 0

Next event: @20 a
Monitoring ended, verdict is: POSITIVE
Monitored 1 events
--------------------------------
$ ./MoniTAal-bin <...> --div a

Monitoring ended, verdict is: POSITIVE
Monitored 0 events
\end{lstlisting}
\caption{Example of monitoring property from \autoref{ex:gt20} ($F_{\ge20}a$) with and without time divergence}
\label{code:past20}
\end{figure}


\subsection{Logic to Automata Translation with \textsc{MightyPPL}}

The tool \mitppl~\cite{hkmmp2025b} was recently developed to generate \acp{tba} from \ac{mtl} with Past and Pnueli modalities (MTLPPL).
The prototype implementation of \monitaal that we present here requires two complementary automata as input, in place of a temporal logic formula. This is general, in that \monitaal is capable of monitoring both safety and liveness properties.
MTLPPL can, with both future and past modalities, similarly also express more than just safety properties.
Note that monitoring a liveness property will never yield the verdict $\bot$, but it is still possible to yield the $\top$ verdict. The property ``$F\ a$'' is a trivial example of this. 

Because \mitppl translates automata to the automata format used in \monitaal we can translate MTLPPL formulas with \mitppl and monitor them with \monitaal.
This allows us to benchmark properties from \timescales~\cite{DBLP:conf/rv/Ulus19} and compare with the monitoring tools \monpoly~\cite{basin2011monpoly} and \reelay~\cite{ulus2026OnlineMonitoringMetric}.

\timescales is a benchmark generator for real-time monitoring tools. It can generate logs (timed words) for 10 safety properties. All of the properties have one upper bounded timing parameter, and some also have a lower bound timing parameter.
For these experiments we monitor each property with 3 different sets of parameters. The lower bounds 3, 30, and 300, and the upper bounds 10, 100, and 1000. The following is a list of the \timescales properties with $l$ and $u$ as the lower and upper bound parameter.

\begin{align*}
\text{AbsentAQ: }& G(q \rightarrow G_{[0, u]} \neg p)\\
\text{AbsentBR: }& G(F_{[0, u]} r \rightarrow (\neg p\ U\ r))\\
\text{AbsentBQR: }& G( (q \land \neg r \land F\ r) \rightarrow (\neg p\ U_{[l,u]}\ r ))\\
\text{AlwaysAQ: }& G(q \rightarrow G_{[0,u]}\ p)\\
\text{AlwaysBR: }& G( F_{0,u}(r) \rightarrow (p\ U\ r) )\\
\text{AlwaysBQR: }& G((q \land \neg r \land F\ r) \rightarrow (p\ U_{[l,u]}\ r))\\
\text{RecurGLB: }& G(F_{[0,u]}  p)\\
\text{RecurBQR: }&  G((q \land \neg r \land F\ r) \rightarrow (F_{[0,u]} (p \lor r)\ U\ r))\\
\text{RespondGLB: }& G( p \rightarrow F_{[l,u]}\ s)\\
\text{RespondBQR: }& G((q \land \neg r \land F\ r) \rightarrow  (p \rightarrow (\neg r\ U_{[l,u]} (\neg r \land s))\ U\ r)\\
\end{align*}
 
For our experiments, we used \timescales to generate logs with a time horizon of 1 million and with an observation on every integer unit of time. The logs generated by \timescales are not exactly 1 million events long for every property. but up to 500 events more in some cases. 
We report on the total running time of the monitor and the peak allocated memory. For \monitaal we also report the time it took to generate the property automata with \mitppl.
We ran each tool on a Linux (Ubuntu) system with an AMD Opteron 6376 CPU. We recorded user time and maximum resident size using the GNU time program\footnote{\url{https://www.gnu.org/software/time/}}. The automata generation time was recorded internally with the C++ Chrono library.

Although we compare \monitaal, \monpoly, and \reelay here, these tools have fundamental differences that make a direct comparison less straightforward than it might seem. 
Both \reelay and \monpoly monitor only safety properties using a finite-word semantics.
\reelay monitors dense-time past-only Metric Temporal Logic~\cite{ulus2026OnlineMonitoringMetric} and  
\monpoly monitors \ac{mfotl} using discrete-time semantics~\cite{basin2008runtime}.
\monitaal with \mitppl monitors MTLPPL (\ac{mtl} with Pnueli, past, and future modalities) and utilizes an infinite word semantics.
\mitppl also implements a slightly different semantics for the Until operator that requires us to extend the formula used by \monitaal to a slightly larger one than that used by the other tools.
As such, a direct comparison of their execution times should be taken with a grain of salt.

Caveats aside, the results in Table~\ref{tab:mightyppl} show that \reelay is the fastest in all cases, while \monpoly is the slowest in all cases. 
The table has a row for each property and a column for the running time (in seconds) of each tool. For \monpoly and \reelay we have no measurement of the initial construction time, but for \monitaal we report the time (in milliseconds) that \mitppl spent on generating the automaton for the property and its negation in the ``Automata'' column. 
The running time shown for \monitaal includes both the time spent on monitoring, as well as automata generation.
For example, the last row shows results for monitoring the property RespondBQR with a lower bound of 300 and upper bound of 1000. For this property (RespondBQR1000), \monitaal 
spent 726 milliseconds on initial automata generation and ran for 2.87 seconds in total while \monpoly spent 6.32 seconds and \reelay spent 0.67 seconds. 

Table~\ref{tab:mightyppl-mem} shows the maximum allocated memory for each monitor in MB.
The table has a row for each property as well as a column for each tool showing their memory usage for each property. 
We see that \reelay uses the most memory in all cases, ranging between 56 and 80 MB, while \monpoly is consistently around 4 MB and \monitaal is between 4 and 22 MB. 
Note that this might not fully represent the memory usage in practical applications of these tools. For example, each monitor is provided a log of data representing a timed word. The required format of this log affects the size of the log file, which might affect the memory usage of the monitor. \reelay's log format is more verbose than the \monpoly format used by the other tools.

It is clear from empirical observation that the memory usage of \monitaal is closely related to the size of the generated automata. AbsentBQR, AlwaysBQR, RespondGLB, and RespondBQR were the only properties for which it took longer than 4 ms to generate the property automata. These are also the only properties for which \monitaal used more than 4 MB of memory. The reason is that the automata for these properties were much larger than the rest.

\begin{table}[ht]
    \begin{tabular}{lrrrr}
      \toprule

    \textbf{Property} & \textbf{Automata} & \textbf{\monitaal} & \textbf{\monpoly} & \textbf{\reelay}\\\midrule
AbsentAQ10        &   2    $\pm$  0   &   1.47  $\pm$  0.01    &     5.32  $\pm$  0.06  &     0.45  $\pm$  0.01\\           
AbsentAQ100       &   2    $\pm$  0   &   1.38  $\pm$  0.01    &     5.26  $\pm$  0.08  &     0.43  $\pm$  0.01\\           
AbsentAQ1000      &   2    $\pm$  0   &   1.38  $\pm$  0.02    &     5.33  $\pm$  0.06  &     0.45  $\pm$  0.02\\\midrule   
AbsentBR10        &   2    $\pm$  0   &   1.67  $\pm$  0.01    &     4.94  $\pm$  0.06  &     0.54  $\pm$  0.01\\           
AbsentBR100       &   2    $\pm$  0   &   1.64  $\pm$  0.02    &     4.94  $\pm$  0.07  &     0.53  $\pm$  0.01\\           
AbsentBR1000      &   2    $\pm$  0   &   1.64  $\pm$  0.02    &     4.95  $\pm$  0.07  &     0.53  $\pm$  0.01\\\midrule   
AbsentBQR10       &   556  $\pm$  5   &   2.96  $\pm$  0.02    &     5.72  $\pm$  0.07  &     0.45  $\pm$  0.01\\           
AbsentBQR100      &   552  $\pm$  10  &   2.22  $\pm$  0.02    &     5.57  $\pm$  0.10  &     0.43  $\pm$  0.01\\           
AbsentBQR1000     &   556  $\pm$  1   &   2.18  $\pm$  0.03    &     5.53  $\pm$  0.08  &     0.43  $\pm$  0.01\\\midrule   
AlwaysAQ10        &   2    $\pm$  0   &   2.11  $\pm$  0.05    &     5.65  $\pm$  0.08  &     0.44  $\pm$  0.01\\           
AlwaysAQ100       &   2    $\pm$  0   &   2.04  $\pm$  0.04    &     5.66  $\pm$  0.07  &     0.42  $\pm$  0.01\\           
AlwaysAQ1000      &   2    $\pm$  0   &   2.03  $\pm$  0.02    &     5.64  $\pm$  0.07  &     0.42  $\pm$  0.01\\\midrule   
AlwaysBR10        &   4    $\pm$  2   &   2.85  $\pm$  0.04    &     5.74  $\pm$  0.07  &     0.55  $\pm$  0.01\\           
AlwaysBR100       &   3    $\pm$  1   &   2.83  $\pm$  0.02    &     5.73  $\pm$  0.04  &     0.54  $\pm$  0.01\\           
AlwaysBR1000      &   3    $\pm$  1   &   2.81  $\pm$  0.02    &     5.82  $\pm$  0.05  &     0.54  $\pm$  0.01\\\midrule   
AlwaysBQR10       &   552  $\pm$  10  &   3.94  $\pm$  0.02    &     6.34  $\pm$  0.08  &     0.46  $\pm$  0.01\\           
AlwaysBQR100      &   555  $\pm$  1   &   3.80  $\pm$  0.03    &     6.36  $\pm$  0.09  &     0.41  $\pm$  0.01\\           
AlwaysBQR1000     &   555  $\pm$  4   &   3.68  $\pm$  0.01    &     6.34  $\pm$  0.10  &     0.40  $\pm$  0.02\\\midrule   
RecurGLB10        &   2    $\pm$  0   &   1.27  $\pm$  0.01    &     1.54  $\pm$  0.03  &     0.62  $\pm$  0.01\\           
RecurGLB100       &   2    $\pm$  0   &   1.11  $\pm$  0.02    &     1.43  $\pm$  0.02  &     0.56  $\pm$  0.01\\           
RecurGLB1000      &   2    $\pm$  0   &   1.09  $\pm$  0.02    &     1.43  $\pm$  0.02  &     0.55  $\pm$  0.02\\\midrule   
RecurBQR10        &   4    $\pm$  0   &   1.39  $\pm$  0.01    &     5.96  $\pm$  0.10  &     0.38  $\pm$  0.01\\           
RecurBQR100       &   4    $\pm$  0   &   1.12  $\pm$  0.01    &     5.89  $\pm$  0.05  &     0.33  $\pm$  0.01\\           
RecurBQR1000      &   4    $\pm$  1   &   1.10  $\pm$  0.01    &     5.84  $\pm$  0.08  &     0.32  $\pm$  0.01\\\midrule   
RespondGLB10      &   559  $\pm$  1   &   4.91  $\pm$  0.01    &     5.47  $\pm$  0.07  &     0.87  $\pm$  0.01\\           
RespondGLB100     &   555  $\pm$  10  &   3.24  $\pm$  0.01    &     5.33  $\pm$  0.11  &     0.79  $\pm$  0.01\\           
RespondGLB1000    &   559  $\pm$  1   &   3.02  $\pm$  0.02    &     5.30  $\pm$  0.08  &     0.77  $\pm$  0.01\\\midrule   
RespondBQR10      &   722  $\pm$  10  &   3.48  $\pm$  0.02    &     6.54  $\pm$  0.11  &     0.76  $\pm$  0.01\\           
RespondBQR100     &   725  $\pm$  1   &   2.93  $\pm$  0.02    &     6.38  $\pm$  0.08  &     0.69  $\pm$  0.01\\           
RespondBQR1000    &   726  $\pm$  1   &   2.87  $\pm$  0.02    &     6.32  $\pm$  0.09  &     0.67  $\pm$  0.01\\\bottomrule

    \end{tabular}
    \caption{Results for running \monitaal (utilizing \mitppl), \monpoly, and \reelay on properties and data from \timescales. For each tool we report the time (in seconds) it took to monitor a word with 1 million events.
    For \monitaal we also report on the time it took to generate the property automata using \mitppl (in milliseconds). All results are the mean values of 10 runs.}
    \label{tab:mightyppl}
  \end{table}

\begin{table}[ht]
    \begin{tabular}{lrrr}
      \toprule

    \textbf{Property} & \textbf{\monitaal} & \textbf{\monpoly} & \textbf{\reelay}\\\midrule

AbsentAQ10        &    4   $\pm$  0        &     4  $\pm$  2         &     56  $\pm$  2 \\           
AbsentAQ100       &    4   $\pm$  0        &     4  $\pm$  2         &     56  $\pm$  2 \\           
AbsentAQ1000      &    4   $\pm$  0        &     4  $\pm$  2         &     56  $\pm$  0 \\\midrule   
AbsentBR10        &    4   $\pm$  1        &     4  $\pm$  2         &     56  $\pm$  2 \\           
AbsentBR100       &    4   $\pm$  0        &     4  $\pm$  2         &     56  $\pm$  2 \\           
AbsentBR1000      &    4   $\pm$  2        &     4  $\pm$  2         &     56  $\pm$  2 \\\midrule   
AbsentBQR10       &    12  $\pm$  3        &     4  $\pm$  2         &     68  $\pm$  2 \\           
AbsentBQR100      &    12  $\pm$  1        &     4  $\pm$  2         &     69  $\pm$  2 \\           
AbsentBQR1000     &    12  $\pm$  1        &     4  $\pm$  2         &     69  $\pm$  0 \\\midrule   
AlwaysAQ10        &    4   $\pm$  2        &     4  $\pm$  2         &     56  $\pm$  2 \\           
AlwaysAQ100       &    4   $\pm$  2        &     4  $\pm$  2         &     56  $\pm$  2 \\           
AlwaysAQ1000      &    4   $\pm$  2        &     4  $\pm$  2         &     56  $\pm$  2 \\\midrule   
AlwaysBR10        &    4   $\pm$  0        &     5  $\pm$  2         &     56  $\pm$  2 \\           
AlwaysBR100       &    4   $\pm$  1        &     5  $\pm$  2         &     56  $\pm$  2 \\           
AlwaysBR1000      &    4   $\pm$  0        &     5  $\pm$  2         &     56  $\pm$  1 \\\midrule   
AlwaysBQR10       &    12  $\pm$  2        &     4  $\pm$  2         &     68  $\pm$  2 \\           
AlwaysBQR100      &    12  $\pm$  2        &     4  $\pm$  2         &     68  $\pm$  2 \\           
AlwaysBQR1000     &    12  $\pm$  2        &     4  $\pm$  2         &     68  $\pm$  2 \\\midrule   
RecurGLB10        &    4   $\pm$  2        &     5  $\pm$  2         &     45  $\pm$  1 \\           
RecurGLB100       &    4   $\pm$  2        &     5  $\pm$  2         &     45  $\pm$  2 \\           
RecurGLB1000      &    4   $\pm$  1        &     5  $\pm$  2         &     45  $\pm$  2 \\\midrule   
RecurBQR10        &    4   $\pm$  2        &     4  $\pm$  2         &     68  $\pm$  1 \\           
RecurBQR100       &    4   $\pm$  2        &     4  $\pm$  2         &     69  $\pm$  2 \\           
RecurBQR1000      &    4   $\pm$  2        &     4  $\pm$  2         &     69  $\pm$  2 \\\midrule   
RespondGLB10      &    12  $\pm$  2        &     4  $\pm$  2         &     57  $\pm$  1 \\           
RespondGLB100     &    12  $\pm$  2        &     4  $\pm$  2         &     57  $\pm$  2 \\           
RespondGLB1000    &    12  $\pm$  2        &     4  $\pm$  2         &     57  $\pm$  2 \\\midrule   
RespondBQR10      &    22  $\pm$  2        &     4  $\pm$  2         &     80  $\pm$  2 \\           
RespondBQR100     &    22  $\pm$  0        &     4  $\pm$  2         &     80  $\pm$  2 \\           
RespondBQR1000    &    22  $\pm$  2        &     4  $\pm$  2         &     80  $\pm$  2 \\\bottomrule

    \end{tabular}
    \caption{Results for memory usage in MB from running \monitaal (utilizing \mitppl), \monpoly, and \reelay on properties and data from \timescales. All results are the mean values of 10 runs.}
    \label{tab:mightyppl-mem}
  \end{table}

\section{Conclusion}
\label{sec:conclusion}

In this work, we have studied the online monitoring problem for timed properties.
We presented an efficient online monitoring algorithm for languages of infinite timed words. We require the language and its complement to be expressed as \acp{tba}, a requirement that is, for example, satisfied for languages specified in the logic \timelogic.
Hence our method is applicable in many realistic scenarios.
In particular, we showed how to account for time divergence which prior work did not readily seem to support.

We also considered an extension of our method supporting timing uncertainty in the time sequence observed by the monitor.
By supporting timing uncertainty in the observed input, we account for real-world systems where the real-valued time-points of events can only be observed up to a given precision.
This limits the soundness of verdicts from monitors that support only concrete timed traces.

The monitoring algorithms for both concrete and uncertain traces has been implemented in the tool \monitaal. The tool 
facilitates intersecting \acp{tba} (also with the time divergence automaton), computing the non-empty language states, reading timed words, and online monitoring.
We aim to exploit the symbolic monitoring engine of \monitaal to replace the current rewrite-based Weighted \ac{mtl} implementation~\cite{bulychev2013rewrite} of \textsc{Uppaal} \ac{smc}~\cite{david2011uppaalsmc}, as this only supports a fragment of \timelogic.

Further improvements to our monitoring algorithm include improved support for uncertainties in the input and additional analysis of timed properties.
In recent work, we introduced support for unobservable symbols in the monitor alphabet~\cite{CGLTZ24}.
This can be logically extended to consider arbitrary mutations to an input sequence modeled in the form of \iac{ta}.


\backmatter

\bmhead{Supplementary information}

The tool implementing the algorithms presented here is freely available at \url{https://github.com/DEIS-Tools/MoniTAal}.

\bmhead{Acknowledgments}

We would like to express our sincere gratitude to Hsi-Ming Ho for helping us integrating the \mitppl\cite{hkmmp2025b} tool.

T.M.\ Grosen, K.G.\ Larsen, and M.\ Zimmermann have been supported by DIREC - Digital Research Centre Denmark.
T.M.\ Grosen has been supported by Mitacs.
S.\ Kauffman has been supported by NSERC - Natural Sciences and Engineering Research Council of Canada.

\section*{Declarations}

The authors have no competing interests to declare that are relevant to the content of this article.

\bibliography{bib}

@inproceedings{DBLP:conf/fct/LarsenPY95,
  author       = {Kim Guldstrand Larsen and
                  Paul Pettersson and
                  Wang Yi},
  editor       = {Horst Reichel},
  title        = {Model-Checking for Real-Time Systems},
  booktitle    = {{FCT} 1995},
  series       = {LNCS},
  volume       = {965},
  pages        = {62--88},
  publisher    = {Springer},
  year         = {1995},
  doi          = {10.1007/3-540-60249-6\_41},
  address={\mbox{}},
  bibsource    = {dblp computer science bibliography, https://dblp.org}
}

@inproceedings{moni,
  author       = {Thomas M{\o}ller Grosen and
                  Sean Kauffman and
                  Kim G. Larsen and
                  Martin Zimmermann},
  editor       = {Patricia Bouyer and
                  Jaco van de Pol},
  title        = {Time for Timed Monitorability},
  booktitle    = {{CONCUR} 2025},
  series       = {LIPIcs},
  pages        = {19:1--19:20},
  publisher    = {Schloss Dagstuhl - Leibniz-Zentrum f{\"{u}}r Informatik},
  year         = {2025},
  xurl          = {https://doi.org/10.4230/LIPIcs.CONCUR.2025.19},
  doi          = {10.4230/LIPICS.CONCUR.2025.19},
  bibsource    = {dblp computer science bibliography, https://dblp.org},
  address = {\mbox{}}
}

@inproceedings{CGLTZ24,
  author       = {Alessandro Cimatti and
                  Thomas M{\o}ller Grosen and
                  Kim G. Larsen and
                  Stefano Tonetta and
                  Martin Zimmermann},
  editor       = {Alexandre Madeira and
                  Alexander Knapp},
  title        = {Exploiting Assumptions for Effective Monitoring of Real-Time Properties
                  Under Partial Observability},
  booktitle    = {{SEFM} 2024},
  series       = {LNCS},
  volume       = {15280},
  pages        = {70--88},
  publisher    = {Springer},
address = {\mbox{}},
  year         = {2024},
  doi          = {10.1007/978-3-031-77382-2\_5},
  bibsource    = {dblp computer science bibliography, https://dblp.org}
}

@inproceedings{FGLZ24,
  author       = {Martin Fr{\"{a}}nzle and
                  Thomas M{\o}ller Grosen and
                  Kim G. Larsen and
                  Martin Zimmermann},
  editor       = {Nikolai Kosmatov and
                  Laura Kov{\'{a}}cs},
  title        = {Monitoring Real-Time Systems Under Parametric Delay},
  booktitle    = {{IFM} 2024},
  series       = {LNCS},
  volume       = {15234},
  pages        = {194--213},
  publisher    = {Springer},
address = {\mbox{}},
  year         = {2024},
  doi          = {10.1007/978-3-031-76554-4\_11},
  bibsource    = {dblp computer science bibliography, https://dblp.org}
}

@inproceedings{GKLZ24,
  author       = {Thomas M{\o}ller Grosen and
                  Sean Kauffman and
                  Kim Guldstrand Larsen and
                  Martin Zimmermann},
  editor       = {Sergiy Bogomolov and
                  David Parker},
  title        = {Monitoring Timed Properties (Revisited)},
  booktitle    = {{FORMATS} 2022},
  series       = {LNCS},
  volume       = {13465},
  pages        = {43--62},
  publisher    = {Springer},
  address      = {Cham},
  year         = {2022},
  url          = {https://doi.org/10.1007/978-3-031-15839-1\_3},
  bibsource    = {dblp computer science bibliography, https://dblp.org}
}

@article{alur1994tba,
  author       = {Rajeev Alur and
                  David L. Dill},
  title        = {A Theory of Timed Automata},
  journal      = {Theor. Comput. Sci.},
  volume       = {126},
  number       = {2},
  pages        = {183--235},
  year         = {1994},
  url          = {https://doi.org/10.1016/0304-3975(94)90010-8},
  doi          = {10.1016/0304-3975(94)90010-8},
  bibsource    = {dblp computer science bibliography, https://dblp.org}
}

@article{alur1996mitl,
  author       = {Rajeev Alur and
                  Tom{\'{a}}s Feder and
                  Thomas A. Henzinger},
  title        = {The Benefits of Relaxing Punctuality},
  journal      = {J. {ACM}},
  volume       = {43},
  number       = {1},
  pages        = {116--146},
  year         = {1996},
  url          = {https://doi.org/10.1145/227595.227602},
  doi          = {10.1145/227595.227602},
  bibsource    = {dblp computer science bibliography, https://dblp.org}
}

@InProceedings{baldor2013monitoring,
  author       = {Kevin Baldor and
                  Jianwei Niu},
  editor       = {Shaz Qadeer and
                  Serdar Tasiran},
  title        = {Monitoring Dense-Time, Continuous-Semantics, Metric Temporal Logic},
  booktitle    = {{RV} 2012},
  series       = {LNCS},
  volume       = {7687},
  pages        = {245--259},
  publisher    = {Springer},
  address      = {Cham},
  year         = {2012},
  doi          = {10.1007/978-3-642-35632-2\_24},
  bibsource    = {dblp computer science bibliography, https://dblp.org}
}

@inproceedings{basin2011monpoly,
  author       = {David A. Basin and
                  Mat{\'{u}}s Harvan and
                  Felix Klaedtke and
                  Eugen Zalinescu},
  editor       = {Sarfraz Khurshid and
                  Koushik Sen},
  title        = {{MONPOLY:} Monitoring Usage-Control Policies},
  booktitle    = {{RV} 2011},
  series       = {LNCS},
  pages        = {360--364},
  publisher    = {Springer},
  year         = {2011},
  xurl          = {https://doi.org/10.1007/978-3-642-29860-8\_27},
  doi          = {10.1007/978-3-642-29860-8\_27},
  bibsource    = {dblp computer science bibliography, https://dblp.org},
  address = {\mbox{}}
}

@article{basin2012algorithms,
  author       = {David A. Basin and
                  Felix Klaedtke and
                  Eugen Zalinescu},
  title        = {Algorithms for monitoring real-time properties},
  journal      = {Acta Informatica},
  volume       = {55},
  number       = {4},
  pages        = {309--338},
  year         = {2018},
  doi          = {10.1007/S00236-017-0295-4},
  bibsource    = {dblp computer science bibliography, https://dblp.org}
}

@InProceedings{basin2008runtime,
  author       = {David A. Basin and
                  Felix Klaedtke and
                  Samuel M{\"{u}}ller and
                  Birgit Pfitzmann},
  editor       = {Ramesh Hariharan and
                  Madhavan Mukund and
                  V. Vinay},
  title        = {Runtime Monitoring of Metric First-order Temporal Properties},
  booktitle    = {{FSTTCS} 2008},
  series       = {LIPIcs},
  volume       = {2},
  pages        = {49--60},
  publisher    = {Schloss Dagstuhl - Leibniz-Zentrum f{\"{u}}r Informatik},
  address =	{Dagstuhl, Germany},
  year         = {2008},
  doi          = {10.4230/LIPICS.FSTTCS.2008.1740},
  bibsource    = {dblp computer science bibliography, https://dblp.org}
}

@inproceedings{bauer2006monitoring,
  author       = {Andreas Bauer and
                  Martin Leucker and
                  Christian Schallhart},
  editor       = {S. Arun{-}Kumar and
                  Naveen Garg},
  title        = {Monitoring of Real-Time Properties},
  booktitle    = {{FSTTCS} 2006},
  series       = {LNCS},
  volume       = {4337},
  pages        = {260--272},
  publisher    = {Springer},
  address      = {Cham},
  year         = {2006},
  doi          = {10.1007/11944836\_25},
  bibsource    = {dblp computer science bibliography, https://dblp.org}
}

@article{bauer2011runtime,
  author       = {Andreas Bauer and
                  Martin Leucker and
                  Christian Schallhart},
  title        = {Runtime Verification for {LTL} and {TLTL}},
  journal      = {{ACM} Trans. Softw. Eng. Methodol.},
  volume       = {20},
  number       = {4},
  pages        = {14:1--14:64},
  year         = {2011},
  xurl          = {https://doi.org/10.1145/2000799.2000800},
  doi          = {10.1145/2000799.2000800},
  bibsource    = {dblp computer science bibliography, https://dblp.org}
}

@InProceedings{brihaye2017mightyl,
  author       = {Thomas Brihaye and
                  Gilles Geeraerts and
                  Hsi{-}Ming Ho and
                  Benjamin Monmege},
  editor       = {Rupak Majumdar and
                  Viktor Kuncak},
  title        = {{MightyL}: {A} Compositional Translation from {MITL} to Timed Automata},
  booktitle    = {{CAV}
                  2017, Part {I}},
  series       = {LNCS},
  pages        = {421--440},
  publisher    = {Springer},
  year         = {2017},
  xurl          = {https://doi.org/10.1007/978-3-319-63387-9\_21},
  doi          = {10.1007/978-3-319-63387-9\_21},
  bibsource    = {dblp computer science bibliography, https://dblp.org},
  address  = {\mbox{}}
}

@InProceedings{bulychev2012monitor,
  author       = {Peter E. Bulychev and
                  Alexandre David and
                  Kim Guldstrand Larsen and
                  Axel Legay and
                  Guangyuan Li and
                  Danny B{\o}gsted Poulsen and
                  Am{\'{e}}lie Stainer},
  editor       = {Nikolaj S. Bj{\o}rner and
                  Andrei Voronkov},
  title        = {Monitor-Based Statistical Model Checking for Weighted Metric Temporal
                  Logic},
  booktitle    = {LPAR 2018},
  series       = {LNCS},
  volume       = {7180},
  pages        = {168--182},
  publisher    = {Springer},
  address      = {Cham},
  year         = {2012},
  doi          = {10.1007/978-3-642-28717-6\_15},
  bibsource    = {dblp computer science bibliography, https://dblp.org}
}

@InProceedings{bulychev2013rewrite,
  author       = {Peter E. Bulychev and
                  Alexandre David and
                  Kim G. Larsen and
                  Axel Legay and
                  Guangyuan Li and
                  Danny B{\o}gsted Poulsen},
  editor       = {Shaz Qadeer and
                  Serdar Tasiran},
  title        = {Rewrite-Based Statistical Model Checking of {WMTL}},
  booktitle    = {{RV} 2012},
  series       = {LNCS},
  volume       = {7687},
  pages        = {260--275},
  publisher    = {Springer},
  address = {Cham},
  year         = {2012},
  doi          = {10.1007/978-3-642-35632-2\_25},
  bibsource    = {dblp computer science bibliography, https://dblp.org}
}

@InProceedings{chattopadhyay2020verified,
  author       = {Agnishom Chattopadhyay and
                  Konstantinos Mamouras},
  editor       = {Jyotirmoy Deshmukh and
                  Dejan Nickovic},
  title        = {A Verified Online Monitor for Metric Temporal Logic with Quantitative
                  Semantics},
  booktitle    = {{RV} 2020},
  series       = {LNCS},
  volume       = {12399},
  pages        = {383--403},
  publisher    = {Springer},
  address      = {Cham},
  year         = {2020},
  doi          = {10.1007/978-3-030-60508-7\_21},
  bibsource    = {dblp computer science bibliography, https://dblp.org}
}

@InProceedings{david2011uppaalsmc,
  author       = {Alexandre David and
                  Kim G. Larsen and
                  Axel Legay and
                  Marius Mikucionis and
                  Zheng Wang},
  editor       = {Ganesh Gopalakrishnan and
                  Shaz Qadeer},
  title        = {Time for Statistical Model Checking of Real-Time Systems},
  booktitle    = {{CAV} 2011},
  series       = {LNCS},
  volume       = {6806},
  pages        = {349--355},
  publisher    = {Springer},
  address      = {Cham},
  year         = {2011},
  doi          = {10.1007/978-3-642-22110-1\_27},
  bibsource    = {dblp computer science bibliography, https://dblp.org}
}

@Article{dewulf2008robust,
  author       = {Martin De Wulf and
                  Laurent Doyen and
                  Nicolas Markey and
                  Jean{-}Fran{\c{c}}ois Raskin},
  title        = {Robust safety of timed automata},
  journal      = {Formal Methods Syst. Des.},
  volume       = {33},
  number       = {1-3},
  pages        = {45--84},
  year         = {2008},
  xurl          = {https://doi.org/10.1007/s10703-008-0056-7},
  doi          = {10.1007/S10703-008-0056-7},
  bibsource    = {dblp computer science bibliography, https://dblp.org}
}

@InProceedings{finkbeiner2009circuits,
  author       = {Bernd Finkbeiner and
                  Lars Kuhtz},
  editor       = {Saddek Bensalem and
                  Doron A. Peled},
  title        = {Monitor Circuits for {LTL} with Bounded and Unbounded Future},
  booktitle    = {{RV} 2009},
  series       = {LNCS},
  volume       = {5779},
  pages        = {60--75},
  publisher    = {Springer},
  address      = {Cham},
  year         = {2009},
  doi          = {10.1007/978-3-642-04694-0\_5},
  bibsource    = {dblp computer science bibliography, https://dblp.org}
}

@InProceedings{geilen2000tableau,
  author       = {Marc Geilen and
                  Dennis Dams},
  editor       = {Mathai Joseph},
  title        = {An On-the-Fly Tableau Construction for a Real-Time Temporal Logic},
  booktitle    = {{FTRTFT} 2000},
  series       = {LNCS},
  volume       = {1926},
  pages        = {276--290},
  publisher    = {Springer},
  address      = {Cham},
  year         = {2000},
  doi          = {10.1007/3-540-45352-0\_23},
  bibsource    = {dblp computer science bibliography, https://dblp.org}
}

@InProceedings{ho2014online,
  author       = {Hsi{-}Ming Ho and
                  Jo{\"{e}}l Ouaknine and
                  James Worrell},
  editor       = {Borzoo Bonakdarpour and
                  Scott A. Smolka},
  title        = {Online Monitoring of Metric Temporal Logic},
  booktitle    = {{RV} 2014},
  series       = {LNCS},
  volume       = {8734},
  pages        = {178--192},
  publisher    = {Springer},
  address      = {Cham},
  year         = {2014},
  doi          = {10.1007/978-3-319-11164-3\_15},
  bibsource    = {dblp computer science bibliography, https://dblp.org}
}

@Article{jahanian1994time,
  author       = {Farnam Jahanian and
                  Ragunathan Rajkumar and
                  Sitaram C. V. Raju},
  title        = {Runtime Monitoring of Timing Constraints in Distributed Real-Time
                  Systems},
  journal      = {Real Time Syst.},
  volume       = {7},
  number       = {3},
  pages        = {247--273},
  year         = {1994},
  xurl          = {https://doi.org/10.1007/BF01088521},
  doi          = {10.1007/BF01088521},
  bibsource    = {dblp computer science bibliography, https://dblp.org}
}

@article{kauffman2021what,
  author       = {Sean Kauffman and
                  Klaus Havelund and
                  Sebastian Fischmeister},
  title        = {What can we monitor over unreliable channels?},
  journal      = {Int. J. Softw. Tools Technol. Transf.},
  volume       = {23},
  number       = {4},
  pages        = {579--600},
  year         = {2021},
  xurl          = {https://doi.org/10.1007/s10009-021-00625-z},
  doi          = {10.1007/S10009-021-00625-Z},
  bibsource    = {dblp computer science bibliography, https://dblp.org}
}

@inproceedings{kauffman2019monitorability,
  author       = {Sean Kauffman and
                  Klaus Havelund and
                  Sebastian Fischmeister},
  editor       = {Bernd Finkbeiner and
                  Leonardo Mariani},
  title        = {Monitorability over Unreliable Channels},
  booktitle    = {{RV} 2019},
  series       = {LNCS},
  volume       = {11757},
  pages        = {256--272},
  publisher    = {Springer},
  address      = {Cham},
  year         = {2019},
  doi          = {10.1007/978-3-030-32079-9\_15},
  bibsource    = {dblp computer science bibliography, https://dblp.org}
}

@Article{koymans1990mtl,
  author       = {Ron Koymans},
  title        = {Specifying Real-Time Properties with Metric Temporal Logic},
  journal      = {Real Time Syst.},
  volume       = {2},
  number       = {4},
  pages        = {255--299},
  year         = {1990},
  url          = {https://doi.org/10.1007/BF01995674},
  doi          = {10.1007/BF01995674},
  bibsource    = {dblp computer science bibliography, https://dblp.org}
}

@inproceedings{li2017practical,
  author       = {Guangyuan Li and
                  Peter Gj{\o}l Jensen and
                  Kim Guldstrand Larsen and
                  Axel Legay and
                  Danny B{\o}gsted Poulsen},
  editor       = {Hakan Erdogmus and
                  Klaus Havelund},
  title        = {Practical controller synthesis for {MTL}$_{0,\infty}$},
  booktitle    = {{SPIN} 2017},
  pages        = {102--111},
  publisher    = {{ACM}},
address = {New York, NY, USA},
  year         = {2017},
  doi          = {10.1145/3092282.3092303},
  bibsource    = {dblp computer science bibliography, https://dblp.org}
}

@Article{moosbrugger2017r2u2,
  author       = {Patrick Moosbrugger and
                  Kristin Y. Rozier and
                  Johann Schumann},
  title        = {{R2U2:} monitoring and diagnosis of security threats for unmanned
                  aerial systems},
  journal      = {Formal Methods Syst. Des.},
  volume       = {51},
  number       = {1},
  pages        = {31--61},
  year         = {2017},
  url          = {https://doi.org/10.1007/s10703-017-0275-x},
  doi          = {10.1007/S10703-017-0275-X},
  bibsource    = {dblp computer science bibliography, https://dblp.org}
}

@InProceedings{nickovic2010dta,
  author       = {Dejan Nickovic and
                  Nir Piterman},
  editor       = {Krishnendu Chatterjee and
                  Thomas A. Henzinger},
  title        = {From {MTL} to Deterministic Timed Automata},
  booktitle    = {{FORMATS} 2010},
  series       = {LNCS},
  volume       = {6246},
  pages        = {152--167},
  publisher    = {Springer},
  address      = {Cham},
  year         = {2010},
  doi          = {10.1007/978-3-642-15297-9\_13},
  bibsource    = {dblp computer science bibliography, https://dblp.org}
}

@article{pike06note,
  author       = {Lee Pike},
  title        = {A Note on Inconsistent Axioms in {Rushby's} {\myquot{{S}ystematic Formal Verification
                  for Fault-Tolerant Time-Triggered Algorithms}}},
  journal      = {{IEEE} Trans. Software Eng.},
  volume       = {32},
  number       = {5},
  pages        = {347--348},
  year         = {2006},
  doi          = {10.1109/TSE.2006.41},
  bibsource    = {dblp computer science bibliography, https://dblp.org}
}

@inproceedings{pnueli2006psl,
  author       = {Amir Pnueli and
                  Aleksandr Zaks},
  editor       = {Jayadev Misra and
                  Tobias Nipkow and
                  Emil Sekerinski},
  title        = {{PSL} Model Checking and Run-Time Verification Via Testers},
  booktitle    = {{FM} 2006},
  series       = {LNCS},
  volume       = {4085},
  pages        = {573--586},
  publisher    = {Springer},
  address      = {Cham},
  year         = {2006},
  doi          = {10.1007/11813040\_38},
  bibsource    = {dblp computer science bibliography, https://dblp.org}
}

@inproceedings{rosu2005monitoring,
  author       = {Prasanna Thati and
                  Grigore Rosu},
  editor       = {Klaus Havelund and
                  Grigore Rosu},
  title        = {Monitoring Algorithms for Metric Temporal Logic Specifications},
  booktitle    = {RV 2004},
  series       = {ENTCS},
  volume       = {113},
  pages        = {145--162},
  publisher    = {Elsevier},
  address = {\mbox{}},
  year         = {2004},
  doi          = {10.1016/J.ENTCS.2004.01.029},
  bibsource    = {dblp computer science bibliography, https://dblp.org}
}

@ARTICLE{rushby1999systematic,
  author       = {John M. Rushby},
  title        = {Systematic Formal Verification for Fault-Tolerant Time-Triggered Algorithms},
  journal      = {{IEEE} Trans. Software Eng.},
  volume       = {25},
  number       = {5},
  pages        = {651--660},
  year         = {1999},
  xurl          = {https://doi.org/10.1109/32.815324},
  doi          = {10.1109/32.815324},
  bibsource    = {dblp computer science bibliography, https://dblp.org}
}

@InProceedings{ulus2014offline,
  author       = {Dogan Ulus and
                  Thomas Ferr{\`{e}}re and
                  Eugene Asarin and
                  Oded Maler},
  editor       = {Axel Legay and
                  Marius Bozga},
  title        = {Timed Pattern Matching},
  booktitle    = {{FORMATS} 2014},
  series       = {LNCS},
  volume       = {8711},
  pages        = {222--236},
  publisher    = {Springer},
  address      = {Cham},
  year         = {2014},
  doi          = {10.1007/978-3-319-10512-3\_16},
  bibsource    = {dblp computer science bibliography, https://dblp.org}
}

@InProceedings{ulus2016online,
  author       = {Dogan Ulus and
                  Thomas Ferr{\`{e}}re and
                  Eugene Asarin and
                  Oded Maler},
  editor       = {Marsha Chechik and
                  Jean{-}Fran{\c{c}}ois Raskin},
  title        = {Online Timed Pattern Matching Using Derivatives},
  booktitle    = {{TACAS} 2016},
  series       = {LNCS},
  volume       = {9636},
  pages        = {736--751},
  publisher    = {Springer},
  address      = {Cham},
  year         = {2016},
  doi          = {10.1007/978-3-662-49674-9\_47},
  bibsource    = {dblp computer science bibliography, https://dblp.org}
}

@inproceedings{DBLP:conf/ac/BengtssonY03,
author       = {Johan Bengtsson and
                  Wang Yi},
  editor       = {J{\"{o}}rg Desel and
                  Wolfgang Reisig and
                  Grzegorz Rozenberg},
  title        = {Timed Automata: Semantics, Algorithms and Tools},
  booktitle    = {Lectures on Concurrency and Petri Nets, Advances in Petri Nets},
  series       = {LNCS},
  volume       = {3098},
  pages        = {87--124},
  publisher    = {Springer},
  address      = {Cham},
  year         = {2003},
  doi          = {10.1007/978-3-540-27755-2\_3},
  bibsource    = {dblp computer science bibliography, https://dblp.org}
}

@inproceedings{DBLP:conf/rtss/DawsY96,
  author       = {Conrado Daws and
                  Sergio Yovine},
  title        = {Reducing the number of clock variables of timed automata},
  booktitle    = {RTSS 1996},
  pages        = {73--81},
  publisher    = {{IEEE} Computer Society},
  year         = {1996},
  xurl          = {https://doi.org/10.1109/REAL.1996.563702},
  doi          = {10.1109/REAL.1996.563702},
  bibsource    = {dblp computer science bibliography, https://dblp.org},
  address = {\mbox{}}
}

@inbook{bouyer08,
author = {Bouyer, Patricia and Laroussinie, François},
publisher = {John Wiley \& Sons, Ltd},
isbn = {9780470611012},
title = {Model Checking Timed Automata},
booktitle = {Modeling and Verification of Real‐Time Systems},
chapter = {4},
pages = {111-140},
doi = {https://doi.org/10.1002/9780470611012.ch4},
xurl = {https://onlinelibrary.wiley.com/doi/abs/10.1002/9780470611012.ch4},
eprint = {https://onlinelibrary.wiley.com/doi/pdf/10.1002/9780470611012.ch4},
year = {2008},
address = {\mbox{}}
}

@article{gear,
  author       = {Magnus Lindahl and
                  Paul Pettersson and
                  Wang Yi},
  title        = {Formal design and analysis of a gear controller},
  journal      = {Int. J. Softw. Tools Technol. Transf.},
  volume       = {3},
  number       = {3},
  pages        = {353--368},
  year         = {2001},
  url          = {https://doi.org/10.1007/s100090100048},
  doi          = {10.1007/S100090100048},
  bibsource    = {dblp computer science bibliography, https://dblp.org}
}

@inproceedings{amara:hal-05043055,
  author       = {Mouloud Amara and
                  Giovanni Bernardi and
                  Mohammed Aristide Foughali and
                  Adrian Francalanza},
  editor       = {Jonathan Aldrich and
                  Alexandra Silva},
  title        = {A Theory of (Linear-Time) Timed Monitors},
  booktitle    = {{ECOOP} 2025},
  series       = {LIPIcs},
  pages        = {1:1--1:30},
  publisher    = {Schloss Dagstuhl - Leibniz-Zentrum f{\"{u}}r Informatik},
  year         = {2025},
  xurl          = {https://doi.org/10.4230/LIPIcs.ECOOP.2025.1},
  doi          = {10.4230/LIPICS.ECOOP.2025.1},
  bibsource    = {dblp computer science bibliography, https://dblp.org},
  address = {\mbox{}}
}

@article{yannakakis1997EfficientAlgorithmMinimizing,
  author       = {Mihalis Yannakakis and
                  David Lee},
  title        = {An Efficient Algorithm for Minimizing Real-Time Transition Systems},
  journal      = {Formal Methods Syst. Des.},
  volume       = {11},
  number       = {2},
  pages        = {113--136},
  year         = {1997},
  url          = {https://doi.org/10.1023/A:1008621829508},
  doi          = {10.1023/A:1008621829508},
  bibsource    = {dblp computer science bibliography, https://dblp.org}
}

@article{ulus2026OnlineMonitoringMetric,
  author       = {Dogan Ulus},
  title        = {Online Monitoring of Metric Temporal Logic using Sequential Networks},
  journal      = {Log. Methods Comput. Sci.},
  volume       = {22},
  number       = {1},
  year         = {2026},
  url          = {https://doi.org/10.46298/lmcs-22(1:12)2026},
  doi          = {10.46298/LMCS-22(1:12)2026},
  bibsource    = {dblp computer science bibliography, https://dblp.org}
}

@inproceedings{basinAERIALAlmostEventRate,
  author       = {David A. Basin and
                  Srdjan Krstic and
                  Dmitriy Traytel},
  editor       = {Giles Reger and
                  Klaus Havelund},
  title        = {{AERIAL:} Almost Event-Rate Independent Algorithms for Monitoring
                  Metric Regular Properties},
  booktitle    = {RV-CuBES 2017},
  series       = {Kalpa Publications in Computing},
  pages        = {29--36},
  publisher    = {EasyChair},
  year         = {2017},
  xurl          = {https://doi.org/10.29007/bm4c},
  doi          = {10.29007/BM4C},
  bibsource    = {dblp computer science bibliography, https://dblp.org},
  address = {\mbox{}}
}

@inproceedings{ulus2017MontreToolMonitoring,
  author       = {Dogan Ulus},
  editor       = {Rupak Majumdar and
                  Viktor Kuncak},
  title        = {Montre: {A} Tool for Monitoring Timed Regular Expressions},
  booktitle    = {{CAV}
                  2017, Part {I}},
  series       = {LNCS},
  pages        = {329--335},
  publisher    = {Springer},
  year         = {2017},
  xurl          = {https://doi.org/10.1007/978-3-319-63387-9\_16},
  doi          = {10.1007/978-3-319-63387-9\_16},
  bibsource    = {dblp computer science bibliography, https://dblp.org},
  address = {\mbox{}}
}

@inproceedings{nickovic2007AMTPropertyBasedMonitoring,
  author       = {Dejan Nickovic and
                  Oded Maler},
  editor       = {Jean{-}Fran{\c{c}}ois Raskin and
                  P. S. Thiagarajan},
  title        = {{AMT:} {A} Property-Based Monitoring Tool for Analog Systems},
  booktitle    = {{FORMATS} 2007},
  series       = {LNCS},
  pages        = {304--319},
  publisher    = {Springer},
  year         = {2007},
  xurl          = {https://doi.org/10.1007/978-3-540-75454-1\_22},
  doi          = {10.1007/978-3-540-75454-1\_22},
  bibsource    = {dblp computer science bibliography, https://dblp.org},
  address = {\mbox{}}
}

@inproceedings{nickovicRTAMTOnlineRobustness2020,
  author       = {Dejan Nickovic and
                  Tomoya Yamaguchi},
  editor       = {Dang Van Hung and
                  Oleg Sokolsky},
  title        = {{RTAMT:} Online Robustness Monitors from {STL}},
  booktitle    = {{ATVA} 2020},
  series       = {LNCS},
  pages        = {564--571},
  publisher    = {Springer},
  year         = {2020},
  xurl          = {https://doi.org/10.1007/978-3-030-59152-6\_34},
  doi          = {10.1007/978-3-030-59152-6\_34},
  bibsource    = {dblp computer science bibliography, https://dblp.org},
  address = {\mbox{}}
}

@inproceedings{henzinger1992WhatGoodAre,
  author       = {Thomas A. Henzinger and
                  Zohar Manna and
                  Amir Pnueli},
  editor       = {Werner Kuich},
  title        = {What Good Are Digital Clocks?},
  booktitle    = {ICALP92},
  series       = {LNCS},
  pages        = {545--558},
  publisher    = {Springer},
  year         = {1992},
  xurl          = {https://doi.org/10.1007/3-540-55719-9\_103},
  doi          = {10.1007/3-540-55719-9\_103},
  bibsource    = {dblp computer science bibliography, https://dblp.org},
  address = {\mbox{}}
}

@inproceedings{havelund2020FirstOrderTimedRuntime,
  author       = {Klaus Havelund and
                  Doron Peled},
  editor       = {Dang Van Hung and
                  Oleg Sokolsky},
  title        = {First-Order Timed Runtime Verification Using {BDDs}},
  booktitle    = {{ATVA} 2020},
  series       = {LNCS},
  pages        = {3--24},
  publisher    = {Springer},
  year         = {2020},
  xurl          = {https://doi.org/10.1007/978-3-030-59152-6\_1},
  doi          = {10.1007/978-3-030-59152-6\_1},
  bibsource    = {dblp computer science bibliography, https://dblp.org},
  address = {\mbox{}}
}

@book{handbook,
  editor       = {Edmund M. Clarke and
                  Thomas A. Henzinger and
                  Helmut Veith and
                  Roderick Bloem},
  title        = {Handbook of Model Checking},
  publisher    = {Springer},
  year         = {2018},
  xurl          = {https://doi.org/10.1007/978-3-319-10575-8},
  doi          = {10.1007/978-3-319-10575-8},
  isbn         = {978-3-319-10574-1},
  bibsource    = {dblp computer science bibliography, https://dblp.org},
  address = {\mbox{}}
}

@inproceedings{henry2025DistributedMonitoringTimed,
  author       = {L{\'{e}}o Henry and
                  Thierry J{\'{e}}ron and
                  Nicolas Markey and
                  Victor Roussanaly},
  editor       = {Erika {\'{A}}brah{\'{a}}m and
                  Houssam Abbas},
  title        = {Distributed Monitoring of Timed Properties},
  booktitle    = {{RV} 2024},
  series       = {LNCS},
  pages        = {243--261},
  publisher    = {Springer},
  year         = {2024},
  xurl          = {https://doi.org/10.1007/978-3-031-74234-7\_16},
  doi          = {10.1007/978-3-031-74234-7\_16},
  bibsource    = {dblp computer science bibliography, https://dblp.org},
  address = {\mbox{}}
}

@inproceedings{hkmmp2025b,
  author       = {Hsi{-}Ming Ho and
                  Shankara Narayanan Krishna and
                  Khushraj Madnani and
                  Rupak Majumdar and
                  Paritosh K. Pandya},
  editor       = {Sebastian Junges and
                  Guy Katz},
  title        = {{MightyPPL}: Model Checking {MITL} with Past and {Pnueli} Modalities},
  booktitle    = {{TACAS} 2026, Part {I}},
  series       = {LNCS},
  pages        = {457--479},
  publisher    = {Springer},
  year         = {2026},
  xurl          = {https://doi.org/10.1007/978-3-032-22752-2\_24},
  doi          = {10.1007/978-3-032-22752-2\_24},
  bibsource    = {dblp computer science bibliography, https://dblp.org},
  address = {\mbox{}}
}

@inproceedings{DBLP:conf/rv/Ulus19,
  author       = {Dogan Ulus},
  editor       = {Bernd Finkbeiner and
                  Leonardo Mariani},
  title        = {Timescales: {A} Benchmark Generator for {MTL} Monitoring Tools},
  booktitle    = {{RV} 2019},
  series       = {LNCS},
  pages        = {402--412},
  publisher    = {Springer},
  year         = {2019},
  xurl          = {https://doi.org/10.1007/978-3-030-32079-9\_25},
  doi          = {10.1007/978-3-030-32079-9\_25},
  bibsource    = {dblp computer science bibliography, https://dblp.org},
  address = {\mbox{}}
}

\end{document}